\RequirePackage{fix-cm}
\documentclass[twocolumn,epjc3]{svjour3}  
\smartqed  % flush right qed marks, e.g. at end of proof
\journalname{Eur. Phys. J. C}
\pdfoutput=1
\usepackage[english]{babel}
\usepackage{bm}
\usepackage{multirow, amsmath, amssymb, amsfonts, array}
\usepackage{mathtools}
\usepackage{color}
\usepackage{bm,graphicx,cancel}
\usepackage{calc}
\usepackage{enumitem}
\usepackage{mathrsfs}
\usepackage{xspace}
\usepackage{spverbatim}
\usepackage{fancyvrb}
\usepackage{xcolor}
\usepackage{listings}
\usepackage{dirtree}
\usepackage{hyperref}
\usepackage{cite}

\newcommand{\rmd}{\mathrm{d}}

\newcommand{\sigmav}{\langle \sigma v \rangle} 
\newcommand{\maddm}{{\sc MadDM}\xspace} 
\newcommand{\mg}{{\sc MG5\_aMC}\xspace} 
\newcommand{\fr}{{\sc FeynRules}\xspace}
\newcommand{\fa}{{\sc FeynArts}\xspace}
\newcommand{\micromegas}{{\sc micrOMEGAs}\xspace}
\newcommand{\nloct}{{\sc NLOCT}\xspace}
\newcommand{\ml}{{\sc MadLoop}\xspace}
\newcommand{\py}{{\sc Pythia}~8\xspace}
\newcommand{\pppc}{{\sc PPPC4DMID}\xspace}

\newcommand{\eg}{{\it e.g.}\xspace}

\newcommand{\hn}{h}
\newcommand{\Hn}{{H^0}}
\newcommand{\An}{A^0}
\newcommand{\Hp}{{H^\pm}}

\renewcommand{\vec}[1]{\boldsymbol{#1}}

\definecolor{blueline}{RGB}{24,89,169}
\definecolor{greenpatch}{RGB}{121,195,106}
\definecolor{redbrickline}{RGB}{161,29,32}

\lstdefinestyle{python}{
    language=Python,
    numbers=left,
    numberstyle=\scriptsize\ttfamily,
    stepnumber=1,
    basicstyle=\small\ttfamily,
    keywordstyle=\small\ttfamily\color{blueline},
    stringstyle=\small\ttfamily\color{greenpatch},
    commentstyle=\small\ttfamily\color{gray},
    emph={self,__init__},
    emphstyle=\small\ttfamily\color{redbrickline},
    escapechar=`,
    frame=l
}

\def\slashb#1{\setbox0=\hbox{$#1$}#1\hskip-\wd0\dimen0=5pt\advance
        \dimen0 by-\ht0\advance\dimen0 by\dp0\lower0.5\dimen0\hbox
          to\wd0{\hss\sl/\/\hss}}
\begin{document}

\title{
Indirect dark-matter detection with \maddm v3.2 -- Lines and Loops
}

\author{Chiara Arina\thanksref{CP3,IRMP}
        \and
        Jan Heisig\thanksref{TTK,CP3} 
        \and
        Fabio Maltoni\thanksref{CP3,Unibo,INFNBO} 
        \and
        Daniele Massaro\thanksref{CP3,Unibo,INFNBO} 
        \and
        Olivier Mattelaer\thanksref{CP3} 
}

\date{}

\institute{Centre for Cosmology, Particle Physics and Phenomenology,  Universit\'e catholique de Louvain, Chemin du Cyclotron~2, B-1348 Louvain-la-Neuve, Belgium \label{CP3}
           \and
           Research Institute for Mathematics and Physics, Universit\'e catholique de Louvain, Chemin du Cyclotron 2, B-1348 Louvain-la-Neuve, Belgium \label{IRMP}
           \and
           Institute for Theoretical Particle Physics and Cosmology, RWTH Aachen University, Sommerfeldstr. 16, D-52056 Aachen, Germany \label{TTK}
           \and
            Dipartimento di Fisica e Astronomia, Alma Mater Studiorum - Universit\`a di Bologna, via Irnerio 46, 40126 Bologna, Italy \label{Unibo}
           \and
           INFN, Sezione di Bologna, viale Berti Pichat 6/2, 40127 Bologna, Italy \label{INFNBO}
}

\maketitle

\begin{abstract}
Automated tools for the computation of particle physics' processes have become the backbone of phenomenological studies beyond the standard model. Here, we present \maddm v3.2. This release enables the fully automated computation of loop-induced dark-matter annihilation processes, relevant for indirect detection observables. 
Special emphasis lies on the annihilation into $\gamma X$, where $X=\gamma, Z, h$ or any new particle even under the dark symmetry. These processes lead to the sharp spectral feature of monochromatic gamma lines -- a smoking-gun signature of dark matter in our Galaxy. \maddm provides the predictions for the respective fluxes near-Earth and derives constraints from the gamma-ray line searches by Fermi-LAT and HESS. As an application, we discuss the implications for the viable parameter space of a top-philic $t$-channel mediator model and the inert doublet model.
\end{abstract}

\section{Introduction}
Indirect detection is one of the three main strategies to search for dark-matter interactions beyond the gravitational one. It explores traces of dark-matter annihilation or decay in cosmic messengers such as photons, neutrinos, positrons, anti-protons or heavier anti-nuclei. While primary annihilation into unstable standard-model particles typically leads to a continuum spectrum of messenger particles, direct annihilation into neutral messengers can give rise to prominent spectral features, such as monochromatic lines~\cite{Bouquet:1989sr,Bergstrom:1989jr,Rudaz:1989ij}. This is a smoking-gun signal for dark matter as the pronounced peak is hardly mimicked by any astrophysical background. While neutrino monochromatic lines typically arise at tree level, see \eg~\cite{Lindner:2010rr,Farzan:2011ck,Arina:2015zoa,ElAisati:2017ppn}, annihilation into photons is a loop-induced process as dark matter is commonly considered to be electrically neutral.
Even though the signal is suppressed by a loop factor, the experimental sensitivity to such sharp energy spectra is high, such that current gamma-ray telescopes, like Fermi-LAT~\cite{Ackermann:2015lka} and HESS~\cite{Abdallah:2018qtu,Abdalla:2018mve}, set strong constraints on the dark-matter parameter space.

Corresponding to the lowest order in perturbation theory, in renormalizable models, the leading diagrams for annihilating into pairs of photons (or of a photon and a neutral particle such as $Z$ and $h$) are non-divergent. Nevertheless, their computation can be involved.
For instance, the number of diagrams to be computed increases considerably for intricate models that contain many (charged) particles. This calls for the need for an automatized procedure to obtain predictions for gamma-ray lines within generic dark matter models.

In this paper, we present \maddm v3.2. It features the automatized computation of loop-induced processes for indirect detection within any dark-matter model for which a UFO~\cite{Degrande:2011ua} model at next-to-leading order (NLO) can be generated. This new module builds upon the indirect detection module, released with \maddm v3.0~\cite{Ambrogi:2018jqj} (see also~Ref.~\cite{Arina:2020kko} for a recent update including a short user guide). At the time of writing, it is the only numerical tool with such capability.  
The currently available public packages provide the computation of loop-induced annihilation cross section for specific models and operators, partly based on analytic expressions obtained in the literature. For instance, \micromegas~\cite{Belanger:2013oya,Belanger:2018mqt} and \textsc{DarkSUSY}~\cite{Gondolo:2004sc,Bringmann:2018lay} include expressions for the supersymmetric neutralino dark matter in the minimal supersymmetric standard model \linebreak(MSSM)~\cite{Bergstrom:1997fh,Bern:1997ng},
and the inert doublet model (IDM)~\cite{Gustafsson:2007pc}. In addition, \micromegas provides automatized computations in several extensions of the MSSM by being linked to the package \textsc{SloopS}~\cite{Baro:2008bg,Baro:2009gn}.
Further results have been obtained for simplified models~\cite{Choi:2012ap,Weiner:2012cb,Tulin:2012uq,Giacchino:2014moa,Ibarra:2014qma}, the next-to-MSSM~\cite{Chalons:2011ia,Chalons:2012xf,Cerdeno:2015jca}, and Kaluza-Klein dark matter~\cite{Jackson:2009kg,Bertone:2009cb,Arina:2014fna}. Alternatively, there have been efforts in obtaining analytic expression in rather model-independent approaches, see \eg~\cite{Ibarra:2014vya,Garcia-Cely:2016hsk} or utilizing an effective theory approach, where the heavy scale physics which gives rise to the interaction between dark matter and the photon is integrated out and the loop processes are reduced to an effective vertex, see~\eg~\cite{Goodman:2010qn,Abazajian:2011tk,Chu:2012qy,Wang:2012ts,Bai:2012qy,Rajaraman:2012db,Rajaraman:2012fu,Dudas:2014ixa,Coogan:2015xla,Duerr:2015vna,Fichet:2016clq}.

Within \maddm~v3.2 we utilize \ml~\cite{Hirschi:2011pa} for the computation of loop-induced annihilation processes. \ml has been embedded in \mg~\cite{Alwall:2011uj} and has been extensively used to compute QCD and EW corrections in the standard model, and QCD corrections \eg~\cite{Backovic:2015soa,Arina:2016cqj,Das:2016pbk,Arina:2020tuw}, and loop-induced processes~\cite{Hirschi:2015iia,Mattelaer:2015haa} in the framework of collider searches for dark matter. The automated computation of cross sections for gamma-lines within \maddm~v3.2 has already been introduced in~\cite{Arina:2020udz} for $t$-channel simplified models, a category of models where loop-induced processes are highly relevant for indirect dark-matter searches, see \eg~\cite{Giacchino:2014moa,Ibarra:2014qma,Arina:2020tuw}. Together with this paper, we release \maddm v3.2 (hereafter simply \maddm). It includes the automated computation of the respective fluxes for a variety of dark-matter density profiles and exclusion limits from utilizing results from the Fermi-LAT satellite and the HESS telescope.

While the focus of this new \maddm release is on sharp spectral features, its capability is more general and accounts for {\it any} loop-induced dark-matter annihilation channel: an example is annihilation into  pair of gluons, which arises at one loop in case of top-philic dark matter, see \eg~\cite{Arina:2016cqj,Garny:2018icg,Colucci:2018vxz}. These processes lead to a continuum gamma-ray energy spectrum because of the showering and hadronization of the gluons and can, for instance, be constrained by the Fermi-LAT dwarf spheroidal analysis~\cite{Fermi-LAT:2016uux,Geringer-Sameth:2014yza}. The new \maddm is designed to recognize the type of energy spectrum produced by the loop-induced process and automatically assign it to the correct analysis pipeline.
Note that the automatized loop process computation is performed exclusively for indirect detection. The automated inclusion of loop-induced processes and higher-order corrections in the relic density computation\footnote{See~refs.~\cite{Baro:2007em,Hellmann:2013jxa,Harz:2016dql,Beneke:2016ync} for related work.} goes beyond the scope of this article and is left for future work.

The remainder of the paper is organised as follows. In section~\ref{sec:loopmaddm}, we discuss the generation of one-loop processes within \maddm, provide details on the required UFO model files and outline the main functionalities of the program. Relevant astrophysical aspects of gamma-ray line searches are detailed in~\ref{sec:aa}, while section~\ref{sec:valid} provides a phenomenological study of gamma-ray line signatures within two simple dark-matter models: a top-philic $t$-channel mediator model and the IDM\@.
We conclude in section~\ref{sec:concl}. 
Further details and technicalities can be found in the appendices. In~appendix~\ref{sec:app0}, we provide a short user guide focusing on the new features. In~appendix~\ref{sec:app1}, we describe the implementation of automatic $J$-factor computation and provide more information about the experimental data. In~appendix~\ref{sec:app3} we describe the treatment of multiple line signatures within \maddm.

\section{Loop-induced processes in \maddm}\label{sec:loopmaddm}

\subsection{MadLoop interface}

One-loop computation is now a very mature topic with a large number of tools and libraries designed for such computations (see the review \cite{Degrande:2018neu} and references therein). In the context of this work, we use \ml \cite{Hirschi:2011pa}, more precisely, the loop-induced implementation of it~\cite{Hirschi:2015iia}. 
The task of \ml is to generate all the Feynman diagrams and to numerically evaluate the numerator of such diagram either as a complex number or as a polynomial in the loop-momenta. Such information can be used by \textsc{Collier}~\cite{Denner:2016kdg}, \textsc{Ninja}~\cite{Mastrolia:2012bu,Peraro:2014cba,Hirschi:2016mdz},  \textsc{CutTools}~\cite{Ossola:2007ax} or \textsc{IREGI}~\cite{Alwall:2014hca}, to decompose the loop in a sum of scalar integrals either employing tensor integral reduction (TIR, introduced by Passarino \& Veltman \cite{Passarino:1978jh}) or by performing the reduction at the integrand level (OPP method \cite{Ossola:2006us}).
Finally, the library OneLoop~\cite{vanHameren:2010cp} or QCDLoop \cite{Carrazza:2016gav} are used to return the finite part and the pole of the associated loop.

A key characteristic of \ml is its flexibility to use all the above-mentioned alternative tools to find the most suitable for the decomposition of a given loop integral. 
Such a decomposition can be numerically unstable (for example in TIR when the Gram determinant vanishes) and can lead to unreliable results. For a given computational tool, \ml assesses the numerical stability of the result either using the estimator of the associated method or by simply re-evaluating the same quantity after a boost or rotation. If the error is higher than a user-specified threshold, another tool is used. If none of the methods encoded in \ml returns a result precise enough, the code redoes the computation in quadruple precision  (available for \textsc{Ninja} and \textsc{CutTools}) which is two orders of magnitude slower. The capacity of \ml to test various reduction algorithms allows minimizing the number of times quadruple precision is needed.

In general, loop computations require special attention to properly handle the finite parts coming from dimensional regularization (where all quantities are defined in $d$ dimension with $d = 4-2\epsilon$). Such terms have two origins. They arise from the epsilon part of the denominator (called $R_1$) and of the numerator (called $R_2$).
Thanks to the fact that the denominators have a simple and well-defined analytical structure, $R_1$ can always be systematically reconstructed in the loop evaluation procedure.
The computation of $R_2$ is more complex and is typically done analytically before the computation of the loop, \emph{i.e.}~at the stage of generating the UFO model files, as described below.

The computation of loop-induced processes is often more challenging than the one related to NLO processes associated with a Born amplitude $B$. This has multiple reasons.
First, in standard computations, one can do the reduction of the loop directly on the amplitude summed/averaged over the helicities of the initial/final states,
while in the case of loop-induced processes, the reduction needs to be performed for each helicity combination independently. To mitigate the slowing down, the strategy used in \mg and \maddm is to Monte-Carlo over helicity contributions. 
Second, in computations at NLO one can avoid evaluating the loop contribution for each phase-space point.
The trick is to use the Born amplitude (up to a normalization factor) as an approximation of the loop amplitude $L$, which corresponds at the integral level at the trivial formula  $\int L = \int \alpha B +\int  (L-\alpha B)$. The application of importance sampling on the latter equality reduces considerably the number of times the loop is computed.
This trick is obviously not possible for loop-induced processes rendering their evaluation computationally intensive. A dedicated effort on the phase-space parametrization (and on the parallelization) has been achieved in \cite{Hirschi:2015iia}. 

To be able to perform loop computations within \mg and \maddm it is necessary to import NLO UFO model files. This can be achieved by using \fr~\cite{Alloul:2013bka}, \fa~\cite{Hahn:2000kx} and \nloct~ \cite{Degrande:2014vpa}. \nloct is a specific software enabling the analytical computation of the $R_2$ terms and ultra-violet (UV) counter-terms for a given model. The latter are required for NLO computation and need to be included in the NLO UFO format, although they are not used for the computation of loop-induced processes. To be more specific, the user has to implement the model in \fr by writing the Lagrangian in the appropriate form. Afterwards, it is renormalized within \fr. To enable electroweak loops the flag \verb!QCDonly! needs to be set to \verb!False!. Feynman gauge should be used.
The renormalized Lagrangian is then passed to \fa, which expresses the various interaction vertices in terms of their couplings and Lorentz structures. 
Subsequently, \nloct is called to solve the renormalization conditions and compute the UV counter-terms and the $R_2$ terms. The lists of the $R_2$ terms and UV counter-terms are written on an external file by \nloct, which must be imported in \fr to obtain the NLO UFO format of the model. 
Depending on the model, the file \verb!.nlo! produced by \nloct could be sizeable, making the exporting of the model slow or unstable.
It is advisable to make use of the \verb!Assumptions! list when running \nloct to specify the possible relations between the parameters of the model (for example by writing explicitly the relations between the particles of the standard model).
Those relations will be used during the computation to shorten the analytical expressions of the UV and $R_2$ terms.
It is possible to export a model using either the $\mathrm{MS}$ or $\overline{\mathrm{MS}}$ renormalization schemes; the latter is recommended. Note that in the presence of unstable internal particles in the loop the use of the complex mass scheme can improve numerical stability. It can be enabled by choosing \verb!ComplexMass->True! in the \verb!WriteCT! command, see~\cite{Degrande:2014vpa} for details.

\subsection{Main functionality}\label{sec:loophowto}

For dark-matter phenomenology, loop-induced annihilation processes with one or two photons in the final state are of particular interest as it provides a smoking-gun signature of monochromatic gamma-lines. 
\maddm automatically generates all contributing diagrams for a given dark-matter model by running the command:
\begin{Verbatim}[fontsize=\fontsize{9.5}{11}\selectfont]
  MadDM> generate indirect_spectral_features
\end{Verbatim}
More precisely, it generates all diagrams for the final states $\gamma\gamma$ and $\gamma X$, where  $X$ includes the $Z$ and $h$ particles of the standard model as well as {\it all} additional beyond standard model (BSM) particles, that are lighter than twice the dark-matter mass and transform even under the dark symmetry that stabilizes dark matter. Individual channels can be generated by explicitly specifying the final state, \emph{e.g.}
\begin{Verbatim}[fontsize=\fontsize{9.5}{11}\selectfont]
  MadDM> generate indirect_spectral_features a z
\end{Verbatim}
Besides the annihilation cross section, \maddm computes the fluxes and the experimental constraints as discussed in section~\ref{sec:aa}. More details on the commands introduced with this release and the corresponding output are provided in~appendix~\ref{sec:app0}.

Loop-induced annihilation into other final states may contribute to the continuum flux of cosmic messengers. A prominent example is annihilation into a pair of gluons, $gg$, that subsequently shower and hadronize. This channel can, for instance, become relevant in top-philic dark-matter models, where dark matter couples only to the top quark at tree level, see \eg~\cite{Arina:2016cqj,Garny:2018icg,Colucci:2018vxz}. 
In the absence of a tree-level diagram for the channel, \maddm automatically switches to the loop-induced mode.
Hence, this annihilation channel is considered by executing the command:
\begin{Verbatim}[fontsize=\fontsize{9.5}{11}\selectfont]
  MadDM> generate indirect_detection g g
\end{Verbatim}
It can also be computed together with tree-level diagrams:
\begin{Verbatim}[fontsize=\fontsize{9.5}{11}\selectfont]
  MadDM> generate indirect_detection
  MadDM> add indirect_detection g g
\end{Verbatim}
Here, the first line leads to the computation of all $2\to2$ tree-level annihilation processes. For more details about the existing syntax of \maddm see~\cite{Ambrogi:2018jqj,Arina:2020kko}.)
After the computation of the annihilation cross section and generation of events, \maddm proceeds with the indirect detection analysis pipeline as introduced in~\cite{Ambrogi:2018jqj}: The energy spectra are either obtained using \py~\cite{Sjostrand:2014zea} or the \pppc tables~\cite{Cirelli:2010xx}. Subsequently, the energy spectra of all contributions (including tree-level contributions if present) are summed up to obtain the total fluxes at source and (if requested) near Earth. The corresponding gamma-ray flux is confronted with constraints from Fermi-LAT observations of dwarf sphe\-roi\-dal galaxies~\cite{Fermi-LAT:2016uux}.

\section{Gamma-ray line phenomenology}\label{sec:aa}

\subsection{Gamma-ray flux}

Photons travel straight (\emph{i.e.}~on geodesics) from the production to the detection point and, hence, they trace their sources.
In general, the differential gamma-ray flux integrated over the region of interest (ROI) in the sky is given by:
\begin{eqnarray}
\frac{{\rmd}\Phi }{\rmd E_\gamma}=   \frac{1}{8 \pi m_\text{DM}^2}\,  \sum_{i} \,\sigmav_i  \frac{{\rmd}N^i_\gamma}{{\rmd}E_\gamma}\,     \int_{\rm ROI} \!{\rmd \Omega}\int_\text{los} \!\rho^2(\vec{r})\,  {\rmd}l ,
\label{eq:difflux}
\end{eqnarray}
where $\sigmav_i$ is the velocity averaged cross-section
of dark-matter particles with a mass $m_\text{DM}$) into final states labeled by $i$. $E_\gamma$ and $N_\gamma$ is the photon energy and the number of photons per annihilation, respectively. Accordingly, ${\rmd} N^i_\gamma /{\rmd}E_\gamma$ is the differential gamma-ray energy spectrum per annihilation. For dark-matter candidates that are not self-conjugated, eq.~\eqref{eq:difflux} has to be multiplied by an additional factor of 1/2.
\maddm provides both the differential flux as well as the total integrated flux. 
The second part of the equation defines the $J$ factor: 
\begin{equation}\label{eq:jfactor}
  J \equiv  \int_{\rm ROI} {\rmd \Omega}\int_\text{los} \rho^2(\vec{r})\,  {\rmd}l\,,
 \end{equation}
 where $\rho(\vec{r})$ denotes the dark-matter density distribution. The second integral integrated is performed over the line of sight (los) $l$.
 The most commonly assumed dark-matter density profiles are spherically symmetric and given by:
\begin{itemize}
    \item Generalised Navarro-Frenk-White (gNFW) \cite{Navarro:1995iw}
    \begin{equation}\label{eq:nfwgen}
    \rho_{\rm gNFW}(r) = \rho_{s} \left( \frac{r_{s}}{r} \right)^{-\gamma} \left({1+\frac{r}{r_{s}}}\right)^{\gamma-3}\,,
    \end{equation}
    %%%
    \item Einasto \cite{Einasto:1965czb}
    \begin{equation}\label{eq:einasto}
    \rho_{\rm Ein}(r) = \rho_{s} \exp \left\{ -\frac{2}{\alpha} \left[\left(\frac{r}{r_{s}}\right)^\alpha -1 \right] \right\}  
    \,,
    \end{equation}
    %%%
    \item Burkert \cite{Burkert:1995yz}
    \begin{equation}\label{eq:burkert}
    \rho_{\rm Burkert}(r) = \rho_{s} \left( 1+\frac{r}{r_{s}}\right)^{-1} \left[1+ \left( \frac{r}{r_s}\right)^2 \right]^{-1} \,,
    \end{equation}
    %%%
    \item Isothermal \cite{2008gady.book.....B}
    \begin{equation}\label{eq:iso}
    \rho_{\rm Iso}(r) = \rho_{s} \left[1 + \left( \frac{r}{r_s}\right)^2  \right]^{-1}\,.
    \end{equation}
\end{itemize} 
In all density profiles, the parameters $r_{s}$ and $\rho_{s}$ are the scale radius and the scale density, respectively. For a given $r_{s}$, $\rho_{s}$ is normalized to match the specified energy density measured at the Sun position (by default $R_\odot = 8.5$ kpc and $\rho_\odot =0.4 \, \rm GeV/cm^3$ is chosen). From the gNFW, eq.~\eqref{eq:nfwgen}, the original NFW and the contracted NFW (NFWc) density profiles are obtained for the choice $\gamma=1$ and $\gamma=1.3$, respectively. For the Einasto profile, $\alpha$ defines the curvature of the density profile, and it is usually fixed at the value $\alpha=0.17$. 
\maddm allows for the computation of the $J$-factors for the above-listed dark-matter density profile with general parameters and a generic parametrization of the ROI, centered around the galactic center. Details about this parametrization and the $J$-factor computation are provided in~appendix~\ref{sec:app1a}. 
   
\subsection{Spectral feature}

Dark matter is assumed to be non-relativistic in galactic halos: typical values within the Milky Way \cite{Eilers:2019} and for similar galaxies are of the order of $10^{-3} c$, while dwarf spheroidal galaxies are expected to host even colder dark matter, $v\sim 10^{-5}c$.
Hence, for annihilation into $\gamma\gamma$ or $\gamma X$, the photon energy spectrum, ${\rmd} N^{\rm line}_\gamma /{\rmd}E_\gamma$ in eq.~\eqref{eq:difflux}, is a sharp spectral line:
\begin{equation}
\label{eq:line_spectrum_delta}
   \frac{\rmd N^{\rm line}_\gamma}{{\rmd}E_\gamma}\,\simeq\,\delta \left(E_\gamma - E_\gamma^{\rm line}\right) \,\times \,\begin{cases} \;2\quad \mbox{for }\;\gamma\gamma \\
   \;1 \quad \mbox{for }\;\gamma X
   \end{cases}
\end{equation}
as a consequence of the kinematics of the two-body annihilation process~\cite{Mateo:1998wg,Weisz:2011gp,Brown:2012uq}. 
Accordingly, 
\begin{equation}
\label{eq:egamma1}
E_\gamma^{\rm line} = m_{\text{DM}} \,,
\end{equation}
for $\gamma\gamma$, and
\begin{equation}
\label{eq:egamma2}
E_\gamma^{\rm line} = m_{\text{DM}} \left(1 - \frac{m^2_X}{4 m^2_{\text{DM}}} \right)\,,
\end{equation}
for $\gamma X$, where $X$ is a
neutral standard-model or BSM particle with mass $m_X$. This characteristic shape allows for peak searches within the experimental data. The implementation of the latter is described in section~\ref{sec:lnlk} together with the specific experimental ROIs for the Fermi-LAT satellite and HESS telescope, optimized for gamma-ray line searches.

\subsection{Experimental constraints}\label{sec:lnlk}

Gamma-ray line signals from dark-matter annihilation are hardly mimicked by any standard astrophysical background. As a consequence, the sensitivity of peak search\-es pursued by the astrophysical experiments is high, typically compensating the loop-suppression of the annihilation rate. As the dark-matter density profile peaks at the Galactic center, it is expected to provide the largest signal. However, depending on the morphology of the continuum gamma-ray background and the cuspiness of the dark-matter density profile, different choices for the ROI maximize the expected sensitivity. Here we consider current upper limits at 95\%  confidence level (CL) on the dark-matter annihilation cross section into a pair of photons from the Fermi-LAT satellite~\cite{Ackermann:2015lka} and the HESS telescope~\cite{Abdallah:2018qtu,Abdalla:2018mve}. Note that gamma-ray line search\-es have also been proposed in dwarf spheroidal galaxies, see \eg~\cite{Lefranc:2016fgn}. However, we do not consider this possibility here.

Fermi-LAT data from the Galactic center, collected over 5.8 years of observation, provide an exclusion bound reaching $ \sigmav_{\gamma \gamma}^{\rm UL} \sim 2 \times 10^{-30} \rm cm^3/s$ for $m_\text{DM}=1$ GeV and $ \sigmav_{\gamma \gamma}^{\rm UL} \sim 3 \times 10^{-28} \rm cm^3/s$ for $m_\text{DM}=500$ GeV for NFWc, the cuspiest dark-matter density profile. 
The analysis takes into account four dark-matter density profiles and associates to each of them an optimized ROI\@. The latter is defined as a circular region centered on the Galactic center, and the galactic plane is masked except for a $12^\circ \times 10^\circ$ box centered on the Galactic center.  The ROIs are R3, R16, R41, and R90, which are optimized for the NFWc (eq.~\eqref{eq:nfwgen} with $\gamma=1.3$), Einasto (eq.~\eqref{eq:einasto}), NFW (eq.~\eqref{eq:nfwgen} with $\gamma=1.0$), and isothermal (eq.~\eqref{eq:iso}) density profiles, respectively.  In the galactic plane, longitudes further than $6^\circ$ from the Galactic center are removed from all ROIs larger than R3, because it is not expected a large dark-matter signal in that region and the photon emission is dominated by standard astrophysical sources. 
Note that the dependence on the dark-matter density profile is sizeable although the use of optimized ROIs mitigates the effect. In the case of isothermal profile (cored profile) the upper bound excludes annihilation cross-section $\sigmav_{\gamma\gamma}^{\rm UL} \sim 8 \times 10^{-29} \, \rm cm^3/s$ for $m_\text{DM}=1$ GeV and $ \sigmav_{\gamma \gamma}^{\rm UL} \sim 3 \times 10^{-27} \rm cm^3/s$ for $m_\text{DM}=500$ GeV. Fermi-LAT data constrain dark-matter masses in the range 200 MeV to 500 GeV.

The data from the center of the Milky Way collected by the HESS telescope concern heavier dark-matter masses, starting from 300 GeV and going up to a maximum of 70 TeV and are based on 254 h of live time observation. The analysis is performed for the Einasto density profile and gives an upper limit of $\sigmav_{\gamma\gamma}^{\rm UL} \sim 4 \times 10^{-28} \, \rm cm^3/s$ at $m_\chi = 500$ GeV and $\sigmav_{\gamma\gamma}^{\rm UL} \sim 2 \times 10^{-25} \, \rm cm^3/s$ at $m_\chi = 70$ TeV. The search has one optimized ROI, called R1, defined as a circular region centered on the Galactic center, with the galactic plane masked at latitudes lower than $0.3^\circ$. The collaboration also quotes the upper limits for a NFW density profile, obtained for R1 simply by rescaling $\sigmav_{\gamma \gamma}^{\rm UL}$ obtained with the Einasto dark-matter density profile by the appropriate $J$-factor. Also for these exclusion limits, the choice of the dark-matter density profile introduces large astrophysical uncertainties. This is known and common to all searches pointing towards the Galactic center, a region where the determination of the spatial distribution of the dark matter is rather uncertain. 

The experimental exclusion limits are provided as a function of ${m_\text{DM}}$ in two forms. First, in terms of the flux $\Phi(E_\gamma)$ and, secondly, translated into annihilation cross section $\sigmav_{\gamma\gamma}$. The latter is computed assuming the $\gamma \gamma$ annihilation channel only. They depend on the chosen dark-matter density profile, \emph{i.e.}~they are anti-proportional to the $J$-factor. We make use of both information 
which is stored within the \texttt{ExpClass} module of \maddm, for further details see~appendix~\ref{sec:app3}. 

The \maddm \texttt{indirect\_spectral\_features} command computes all $2\to2$ processes at one-loop order characterized by at least one photon in the final state, \emph{i.e.}~$\gamma\gamma$ and $\gamma X$, and confronts it with the above-mentioned experimental constraints. 
Two types of results are shown. On the one hand, we display the cross sections for all individual channels together with the respective experimental limits.\footnote{The limits for $\gamma X$ are obtained from the experimental ones for $\gamma\gamma$ by the rescaling $
\sigmav_{\gamma X}^\text{UL}= 2\, (m_\text{DM}/E^\text{line}_\gamma)^2\,\sigmav_{\gamma\gamma}^\text{UL} (E_\gamma^\text{line})$, 
where $E_\gamma^\text{line}$ is the position of the line according to eq.~\eqref{eq:egamma2}.} Note that we neither combine different channels (if their $E_\gamma^\text{line}$ is equal within the experimental resolution) nor check the applicability of the experimental analysis (in case of multiple distinguishable lines; see below) for this type of results.

On the other hand, we  list all spectral lines as they would occur in the experimental analysis and display the respective fluxes together with the corresponding 95\% CL limits for both experiments. To this end, \linebreak\maddm combines all channels whose peaks are sufficiently close in energy such that they are indistinguishable given the experimental resolution. This is, in particular, relevant for heavy dark matter, \emph{i.e.}~for $4m_\text{DM}^2/m_X^2\gg 1$. Note that the energy resolution is experiment-specific.  Furthermore, (combined) spectral lines lacking a sufficient separation from each other  question the applicability of the experimental limit-setting procedure and are flagged accordingly. The minimal separation and combination conditions (in terms of the experimental resolution) can be adjusted by the user. The merging procedure is detailed in~appendix~\ref{sec:app3}.

While \maddm uses the Einasto profile by default, the user can freely choose the dark-matter density profile (among the ones listed in Sec.~\ref{sec:aa}) and its parameters in the file \texttt{maddm\_card.dat}  as well as the ROI for the Fermi-LAT analysis, see~appendix~\ref{sec:app1a} for details. The program automatically computes the corresponding $J$-factors. 
However, note that the cross section upper limits from Fermi-LAT are only displayed for the combination of ROI and profile chosen in the analysis.
Similarly, the display of limits is omitted outside the energy range for which the analyses have been performed. Finally, we would like to mention that the user can easily implement other experiments \emph{e.g.} for the computation of projected limits. A template experiment is provided with the code, see~appendix~\ref{sec:app1b} for details.

On top of the above results, \maddm computes an approximate likelihood from the predicted flux for two of the ROIs of the Fermi-LAT analysis, namely R3 and R16. In Ref.~\cite{Cuoco:2016jqt}, a likelihood profile has been derived from the Fermi-LAT data as a function of the integrated flux in these ROIs. We utilize these results within \maddm.
The likelihood functions extend the \maddm experimental module \texttt{ExpConstraints}.

\section{Physics applications}\label{sec:valid}

To demonstrate the physics impact, in the following we apply the new feature of \maddm\ to two BSM scenarios and derive gamma-ray line constraints on their model parameter space. We consider a simplified top-philic dark-matter model as well as the IDM as illustrative examples.

\subsection{Simplified top-philic dark-matter model}\label{sec:stopmodel}

This scenario is a simplified dark-matter model that supplements the standard model by a top-philic scalar mediator, 
$\tilde t$, and a Majorana fermion, $\chi$. The former has gauge quantum numbers identical to the right-handed top quark, while the latter is 
a singlet under the standard-model gauge interactions. Imposing a $Z_2$ symmetry under which only $\chi$ and $\tilde t$ transform 
oddly, we can stabilize $\chi$ for $m_{\chi}<m_{\tilde t}$, rendering it a viable dark-matter candidate. 

The interactions with the standard model are described by the Lagrangian
\begin{equation}
    \mathcal{L}_\text{int} = |D_\mu \tilde t|^2 + \lambda_\chi \tilde t\, \bar{t}\,\frac{1-\gamma_5}{2}\chi +\text{h.c.}\,,
    \label{eq:stopmodel}
\end{equation}
where $D_\mu$ denotes the gauge covariant derivative, $t$ the top quark Dirac field and $\lambda_\chi$ is the dark-matter coupling. The two masses and $\lambda_\chi$ are considered free parameters of the model.\footnote{For a specific coupling, $\lambda_\chi=2\sqrt{2}e/(3\cos \theta_W)$, eq.~\eqref{eq:stopmodel} resembles the limiting case of the MSSM where only the right-handed stop and a bino-like neutralino are light and, hence, phenomenologically relevant.}

 \begin{figure}[t]
\centering
\includegraphics[trim={2mm 0 10mm 6.5mm}, clip,width=.47\textwidth]{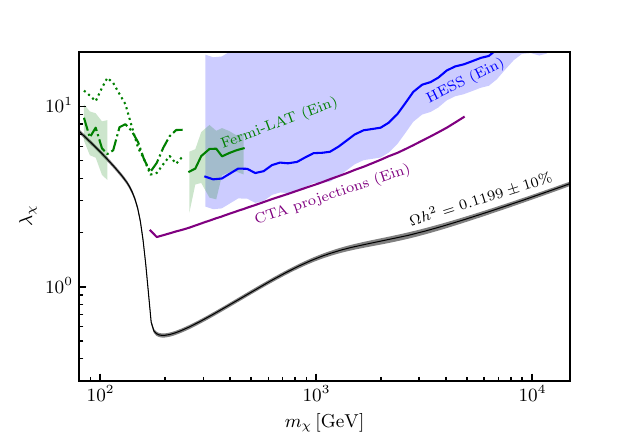}
\caption{Parameter space for the top-philic model, including the constraints for loop-induced dark-matter annihilation into monochromatic photons.
The black solid curve (and shaded band around it) shows the dark-matter coupling that yields the measured relic density, assuming $\Omega h^2=0.1199$ (with a 10\% uncertainty on the theory prediction), taken from~\cite{Garny:2018icg}.
The green (blue) curve denotes the Fermi-LAT (HESS) limits for the Einasto profile. The lower and upper boundaries of the shaded bands indicate the corresponding limits for the NFWc and Isothermal profile, respectively.
The dotted and dot-dashed curves show the upper limits for the individual final states $\gamma\gamma$ and $\gamma Z$, respectively, in the range where the signals are not combined (see text for details).
The purple curve shows projections for CTA assuming the Einasto profile.
}
\label{fig:topphilic}
\end{figure}

Such $t$-\emph{channel mediator} or \emph{charged parent} models have been widely studied in the literature. In particular, the top-philic case considered here has been studied in detail in a large range of $\lambda_\chi$~\cite{Garny:2018icg}. Through thermal freeze-out, the model can explain the relic density in a wide range of masses with perturbative couplings. For the slice in parameter space with $\Delta m = m_{\tilde t}- m_{\chi}=m_t$, shown in figure~\ref{fig:topphilic}, the smallest coupling is required around $m_\chi\gtrsim m_t$ where the annihilation into a pair of top-quarks opens up. For smaller dark-matter masses, this channel becomes Boltzmann suppressed during freeze-out and the three-body and loop-induced annihilation processes $t Wb$ and $gg$, respectively, are dominant~\cite{Garny:2018icg}. 

We generate the NLO UFO model with \fr \cite{Alloul:2013bka}, \fa~\cite{Hahn:2000kx} and \nloct~\cite{Degrande:2014vpa}
and compute the loop-induced annihilation into $\gamma\gamma$, $\gamma Z$ and $\gamma h$ that occur via triangle and box diagrams involving the scalar mediator and the top-quark in the loop. They involve a total of 14, 14, and 6 diagrams, respectively. The computation of the $\gamma\gamma$ contribution has been performed in the literature for the first time in~\cite{Bergstrom:1989jr,Rudaz:1989ij} (see also~\cite{Garny:2013ama,Garny:2018icg}). Our results agree with the ones from~\cite{Bergstrom:1989jr,Garny:2018icg}.

The resulting constraints on the dark-matter coupling, $\lambda_\chi$, are shown in figure~\ref{fig:topphilic}. The green and blue lines denote the limits from Fermi-LAT and HESS, respectively, assuming an Einasto profile. The corresponding shaded bands around them indicate the uncertainty from the density profile. Its lower and upper boundaries mark the limits assuming the NFWc and the isothermal profile, respectively. Notably, the band for the case of Fermi-LAT is considerably smaller as a result of optimized ROIs for the different profiles, see section~\ref{sec:lnlk}. The dot-dashed curve denotes the projection for CTA \cite{Consortium:2010bc} taken from Ref.~\cite{Hryczuk:2019nql}.

For $m_\chi \gtrsim 240$\,GeV, the energy of the spectral lines arising from annihilation into $\gamma h$ and $\gamma Z$ are equal within the detector resolution of Fermi-LAT and hence combined for the limit setting. They are combined with $\gamma \gamma$ above $m_\chi \gtrsim 260$\,GeV. However, the $\gamma h$ annihilation cross section is $p$-wave suppressed and hence negligible (the corresponding limit for the individual channel is far outside the displayed window). The limits arising from combinations of channels are drawn as solid lines.

For $m_\chi \lesssim 260$\,GeV we display the limits for the individual channels; the dotted and dot-dashed lines correspond to $\gamma\gamma$ and $\gamma Z$, respectively. However, in the range $110\,\text{GeV} \lesssim m_\chi \lesssim 260\,\text{GeV}$, the peaks are not sufficiently separated while their integrated fluxes are of similar size questioning the validity of the limit-setting procedure in the Fermi-LAT analysis (see appendix~\ref{sec:app3} for further details). This is indicated by the missing green band which is only shown in the ranges providing a reliable limit. Note that for  $m_\chi \lesssim 110 \,\text{GeV}$ it corresponds to the stronger limit ($\gamma Z$).

Current limits only constrain the region below \linebreak$110\,\text{GeV}$ for the NFWc profile. Note that this region is also challenged by direct detection experiments (see~\cite{Garny:2018icg}) and -- independent on astrophysical parameters -- by LHC searches for supersymmetric top partners~\cite{Sirunyan:2020tyy,Sirunyan:2021mrs,Aad:2020aob,Aad:2021hjy}. However, for the considered slice $\Delta m \simeq m_t$ the signal acceptance is subject to large uncertainties and, hence, this region is often blanked out in the experimental limits, see \emph{e.g.}~\cite{Sirunyan:2020tyy}.

\subsection{Inert doublet model}

The IDM supplements the standard model by an additional (\emph{inert}) Higgs doublet, $\Phi$, that is odd under an exact $Z_2$ symmetry, stabilizing the lightest of its state. The interactions with the standard model arise from the gauge kinetic terms for $\Phi$ and the scalar potential, which reads
\begin{equation}
\begin{split}
	V =& \;\mu_1^2 |H|^2  + \mu_2^2|\Phi|^2 + \lambda_1 |H|^4+ \lambda_2 |\Phi|^4 
	+ \lambda_3 |H|^2| \Phi|^2 \\
		&+ \lambda_4 |H^\dagger\Phi|^2 
		+ \lambda_5/2\,\big[ (H^\dagger\Phi)^2 + \mathrm{h.c.} \big].
\label{Eq:TreePotential}
\end{split}
\end{equation}
After electroweak symmetry breaking, the model contains a total of five physical scalar states with masses given by
\begin{subequations}
\begin{align}
	&m_{\hn}^2 = \mu_1^2 + 3 \lambda_1 v^2\,,\\
	&m_{\Hn}^2= \mu_2^2 + \lambda_L v^2\,, \\
	&m_{\An}^2 = \mu_2^2 + \lambda_S v^2\,,\\
	&m_{\Hp}^2 = \mu_2^2 + \frac{1}{2} \lambda_3 v^2\,, 
\end{align}
\end{subequations}
%
%\begin{equation}
%	m_{\hn}^2 = \mu_1^2 + 3 \lambda_1 v^2\,,\quad
%	m_{\Hn}^2= \mu_2^2 + \lambda_L v^2\,, \quad
%	m_{\An}^2 = \mu_2^2 + \lambda_S v^2\,,\quad
%	m_{\Hp}^2 = \mu_2^2 + \frac{1}{2} \lambda_3 v^2\,, 
%\end{equation}
where
\begin{equation}
	\lambda_{L,S} = \frac{1}{2} \left( \lambda_3 + \lambda_4 \pm \lambda_5 \right)\,.
\end{equation}
Imposing $m_{\hn} \simeq 125$\,GeV, five free model parameters remain. We express them as \{$m_{\Hn}$, $m_{\An}$, $m_{\Hp}$, $\lambda_L$, $\lambda_2$\}. For details about the parameters and the IDM, we refer to~\cite{Barbieri:2006dq,LopezHonorez:2006gr,Honorez:2010re}.

Despite its simplicity, the IDM leads to a rich phenomenology. It provides a viable dark-matter candidate
in various regions in parameter space involving all ``exceptional'' regimes~\cite{Griest:1990kh}  of dark-matter freeze-out -- annihilation near a resonance, close to thresholds and co-annihilation, see, \eg~\cite{Goudelis:2013uca,Ilnicka:2015jba,Belyaev:2016lok,Eiteneuer:2017hoh}. 

Here, we assume $\Hn$ to be the dark-matter candidate and focus on the region around $m_{\Hn} \simeq 72$ GeV, which has been previously studied by some of the authors in~\cite{Eiteneuer:2017hoh}.
In this region, the measured relic density can be explained by annihilation into $WW^*, ZZ^*$ via the gauge kinetic interaction alone~\cite{Honorez:2010re,Banerjee:2021xdp}, where $V^*$ denotes an off-shell vector boson. Accordingly, the $\Hn$-Higgs coupling, $\lambda_L$, can be arbitrarily small.
Interestingly, it is among the two regions that provide a good fit to the Fermi-LAT Galactic center gamma-ray excess when interpreted as a signal of dark matter (the other one being the resonant region which, however, requires highly tuned parameters)~\cite{Eiteneuer:2017hoh}. At the same time, it is unchallenged by current limits from the LHC~\cite{Dercks:2018wch} and direct detection -- due to the small $\lambda_L$, diagrams with massive gauge bosons in the loop~\cite{Klasen:2013btp} dominate the direct detection cross section which are around an order of magnitude smaller than the current limit from LZ~\cite{LZ:2022ufs}.
While upcoming observations of continuum gamma-ray emissions will provide the sensitivity to probe this region~\cite{Eiteneuer:2017hoh}, in the presence of the yet unexplained excesses in the continuum emissions, an independent probe of the model via gamma-line searches is highly desirable. Hence, loop-induced annihilation into $\gamma\gamma$ and $\gamma Z$ is a smoking-gun signature that allows us to test its dark-matter origin.  

Again, we generate the NLO UFO model with \fr~\cite{Alloul:2013bka}, \fa~\cite{Hahn:2000kx} and \nloct~ \cite{Degrande:2014vpa}. The processes $H^0 H^0\to \gamma\gamma$ and $H^0 H^0\to\gamma Z$ 
involve 140 and 172 diagrams, respectively. 
Note that $H^0 H^0\to \gamma h$ is forbidden due to charge-conjugation invariance (\emph{i.e.}~a generalization of Furry's theorem).
We have validated our numerical setup using existing results for $\langle\sigma v\rangle_{\gamma\gamma}$ in the literature~\cite{Gustafsson:2007pc,Garcia-Cely:2016hsk} and found agreement within the numerical precision.\footnote{The results exhibit a noticeable dependence on the choice of standard-model parameters of the electroweak sector. This is caused, on the one hand, by the mapping of different sets of IDM input parameters into each other and, on the other hand, by the evaluation of the cross section itself.}

We consider points within the $2\sigma$ region from the parameter scan and fit performed in~\cite{Eiteneuer:2017hoh}, which takes into account constraints from the relic density~\cite{Ade:2015xua},
electroweak precision observables~\cite{Baak:2014ora,Eriksson:2009ws}, new physics searches at LEP-II~\cite{Pierce:2007ut,Lundstrom:2008ai}, 
indirect detection searches for continuum $\gamma$-ray spectra from dwarf spheroidal galaxies~\cite{Fermi-LAT:2016uux}
and theoretical requirements of unitarity, perturbativity and vacuum stability computed with \linebreak\textsc{2HDMC}~\cite{Eriksson:2009ws}.
The scan was performed with \textsc{Multinest}~\cite{Feroz:2008xx,Feroz:2013hea} for efficient parameter sampling. 

\begin{figure*}[t!]
\centering
\includegraphics{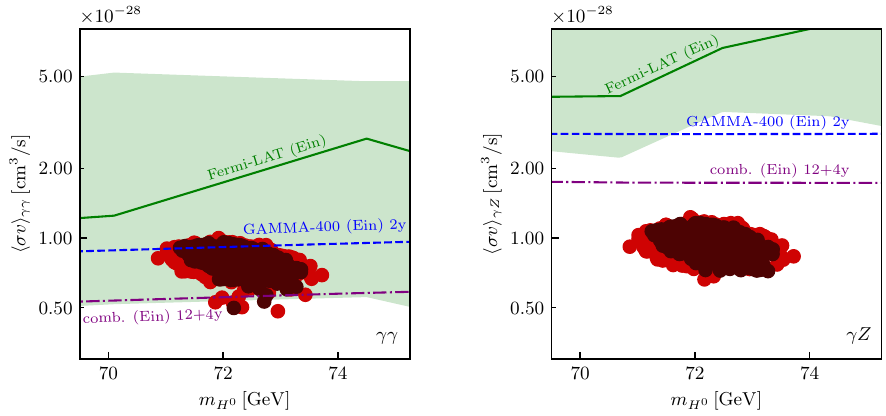}
\caption{Cross section for the loop-induced dark-matter annihilation into $\gamma\gamma$ (left panel) and $\gamma Z$ (right panel) for the allowed IDM parameter points explaining all of dark matter ($\Omega h^2=0.12$) in the range $70\sim75\,$GeV. Parameter points in the 1 and $2\sigma$ regions are drawn in dark brown and red, respectively.
The green solid curve denotes the Fermi-LAT limit for the Einasto profile. The lower and upper boundary of the green shaded band shows the respective limits for NFWc and isothermal profile, respectively. The blue dotted and purple dot-dashed curves show projections for GAMMA-400 (2 years) and the combination of Fermi-LAT (12 years) and GAMMA-400 (4 years), respectively, both assuming the Einasto profile.}
\label{fig:IDprospIDM}
\end{figure*}

The resulting cross sections for $\langle\sigma v\rangle_{\gamma\gamma}$ and $\langle\sigma v\rangle_{\gamma Z}$ are shown in figure~\ref{fig:IDprospIDM} together with the corresponding limits from Fermi-LAT~\cite{Ackermann:2015lka} and future projections for GAMMA-400 (2 years) as well as a combination of both observations (with 12 and 4 years observational time, respectively)~\cite{Egorov:2020cmx}. All lines assume the Einasto profile while the shaded band around the Fermi-LAT limit illustrates the uncertainty due to the choice of  the profile. The upper and lower boundary of the shaded band corresponds to the limit assuming the isothermal and NFWc profile, respectively. In the $\gamma\gamma$ channel, the current limits only constrain the considered region  for the case of NFWc profile. For the Einasto profile, future observations are expected to provide sensitivity. In the case of an excess in $\gamma\gamma$, the confirmation of a line-signal around $E_\gamma \simeq43\,$GeV in $\gamma Z$ would be important to establish the model. However, the sensitivity of the combination of Fermi-LAT and GAMMA-400 (purple, dot-dashed curve in right plot) is still too low by about a factor of two to reach the expected signal originating from $\sigmav_{\gamma Z}$.

%=====================
%    \                                           |
%      \                                         |
%        \                                       |
%                                      \         |
%                                        \       |
%                                          \     |
%=====================

\section{Discussion and conclusions}\label{sec:concl}

Gamma-ray lines are smoking-gun signatures of direct dark-matter annihilation into photons in galactic halos. Additionally, they can complement searches for continuum gamma-ray emission from dark matter. This is, in particular, relevant given the presence of unexplained excesses in the continuum gamma-ray flux, like the long-standing Fermi-LAT Galactic center excess. 

For electrically neutral dark matter, annihilation into $\gamma\gamma$ or $\gamma X$ is loop-induced. While the respective cross sections have been computed for a variety of models, with this paper, we present \maddm v3.2 that enables their automated  computation for arbitrary dark-matter models implemented in the NLO UFO format. It utilizes an interface to \ml, a \mg tool for computing loop processes,
optimized for the involved kinematics of non-relativistic dark-matter annihilation. Furthermore, \maddm computes the integrated fluxes and applies experimental constraints from Fermi-LAT and HESS to the model. Allowing for the computation of the $J$-factor for a variety of dark-matter density profiles as well as regions of interest, it provides the user with the necessary flexibility for comprehensive dark-matter studies. 

We demonstrated the capabilities of the program by applying it to two relevant physics scenarios. First, we studied the constraints from line-signatures on a simplified top-philic dark-matter model. It provides a viable dark-matter candidate over a large range of masses. For dark matter lighter than the top-quark, $2\to2$ annihilation processes at tree-level are suppressed, while loop-induced processes become important and consequently require large dark-matter couplings to match the relic density. This also enhances the strength of the gamma-line signals relative to the continuum emission for indirect detection. However, current line searches from Fermi-LAT only constrain the region of dark-matter masses below roughly 110\,GeV. 

According to the given experimental energy resolution, \maddm merges the spectral feature of various channels improving the constraining power. For instance, for the channels $\gamma\gamma$ and $\gamma Z$, this happens above $m_\text{DM} = 260\,\text{GeV}$ in the case of the Fermi-LAT constraints, while their signals could be merged for the entire reported range for the current HESS and projected CTA limits. Nevertheless, the model remains unchallenged by these limits.

As a second scenario, we considered the inert doublet model. While it provides several viable regions in the parameter space, we concentrate on the region \linebreak$71\,\text{GeV}\lesssim m_\text{DM} \lesssim 74\,\text{GeV}$ that yields the measured relic density without any particular tuning of the involved parameters, providing the required annihilation rate via annihilation into $WW^*, ZZ^*$ due to the stand\-ard-model gauge interactions. Furthermore, this region fits the Fermi-LAT Galactic center excess and is currently not challenged by direct detection or collider searches. We found that current limits from gamma-line observation constrain the scenario for the very cuspy NFWc profile while future observations by GAMMA-400 and Fermi-LAT are expected to probe it for a slightly broader range of dark-matter profiles. An intriguing possibility is the observation of two peaks, from $\gamma\gamma$ and $\gamma Z$ which are well separated in the considered parameter region. However, the $\gamma Z$ signal still appears out-of-reach according to the above-mentioned projections.

While the main focus of this paper is on gamma-line signatures, we would like to stress that the current release of \maddm allows for the computation of {\it arbitrary} (one-)loop-induced dark-matter annihilation processes, \emph{i.e.}~it is not restricted to photon final states. Annihilation into a pair of gluons is a relevant example that falls into this class. Unlike annihilation into photons, it leads to a continuum spectrum which is confronted with experimental constraints that we introduced in an earlier release of \maddm.

This release can be seen as the first step towards incorporating automated loop-level computations in \linebreak\maddm. In the light of current data from the cosmic microwave background enabling a relic density measurement below percent level precision, higher-order corrections in the corresponding theoretical predictions become relevant and call for automated computational tools. While \mg utilizes a suitable framework for this quest, an efficient computation of thermally averaged cross sections at NLO requires further research that is left for future work. Another planned development of \maddm concerns generic one-loop computations of direct detection cross sections.  They are of phenomenological relevance for a variety of dark-matter models. In fact, in the two physics scenarios considered in this paper, the leading contribution to dark-matter scattering off nuclei is loop-induced.

\section*{Acknowledgements}
We thank C\'eline Degrande for support with \textsc{NLOCT}, Alessandro Cuoco for welcomed input regarding the Fermi-LAT likelihood function and Valentin Hirschi for fruitful discussions on the optimization of \ml within \maddm. Computational resources have been provided by the supercomputing facilities of the Université catholique de Louvain (CISM/UCL) and the Consortium des \'Equipements de Calcul Intensif en \linebreak F\'ed\'eration Wallonie Bruxelles (C\'ECI) funded by the Fond de la Recherche Scientifique de Belgique (F.R.S.-FNRS) under convention 2.5020.11 and by the Walloon Region. This work has received funding from the European Union's Horizon 2020 research and innovation programme as part of the Marie Skłodowska-Curie Innovative Training Network MCnetITN3 (grant agreement no.~722104). CA and DM have been supported by the Innoviris  ATTRACT 2018 104 BECAP 2 agreement.
JH acknowledges support from the F.R.S.-FNRS as a Charg\'e de recherche and the DFG via the Collaborative Research Center TRR 257.
\appendix

\section{Getting started}\label{sec:app0}

In this appendix we introduce the new features and main modifications of the code. For further information on the existing functionality, we refer the reader to the short user guide provided in~\cite{Arina:2020kko} as well as the publication accompanying the release of v3.0~\cite{Ambrogi:2018jqj}.

\subsection{Installation}

The current version of \maddm can be downloaded and installed directly as a \mg plug-in. To this end, the user downloads and untars the latest stable version of \mg from launchpad (\url{https://launchpad.net/mg5amcnlo/+download}). At the time of writing, the latest long-term stable version is 2.9.4, which we assume for definiteness in the following displayed examples. Note that \maddm currently requires Python~2.7, while \mg, is compatible with both Python 2 and 3.\footnote{Note that from version 2.8 on,  \mg requires the Python module \texttt{six}.} Additionally, the \texttt{SciPy} and \texttt{NumPy} modules have to be installed. 
Once \mg has been untared, the user can enter the corresponding directory, start \mg and install \maddm\ via the command line:
\begin{Verbatim}[fontsize=\fontsize{9.5}{11}\selectfont]
  mydir$ tar -xzf MG5_aMC_v2.9.4.tar.gz
  mydir$ cd MG5_aMC_v2_9_4/
  MG5_aMC_v2_9_4$ python2.7 bin/mg5_aMC
  MG5_aMC> install maddm
  MG5_aMC> quit
  MG5_aMC_v2_9_4$
\end{Verbatim}
The \maddm source code is automatically copied to \texttt{MG5\_aMC\_v2\_9\_4/PLUGIN/maddm}  while the executable \linebreak python file \texttt{maddm.py} is stored in \texttt{MG5\_aMC\_v2\_9\_4/bin/}. 
Alternatively, the \maddm package is also available from launchpad (\url{https://launchpad.net/maddm}).

As described in section~\ref{sec:loopmaddm}, \ml requires several additional tools, utilized for the evaluation of loop integrals, namely \textsc{Ninja}~\cite{Mastrolia:2012bu,Peraro:2014cba,Hirschi:2016mdz}, \textsc{Collier}\cite{Denner:2016kdg}, \textsc{CutTools}~\cite{Ossola:2007ax} and \textsc{IREGI}~\cite{Alwall:2014hca}. When computing a loop-induced process for the first time, \maddm asks the user to confirm their automatic installation within the launch interface (see below).\footnote{Note that the installation of \textsc{Collier} requires \texttt{cmake}. If \texttt{cmake} is not installed, the installation of \textsc{Collier} is not enabled by default. We recommend enabling the installation following the command line output. Note that this creates a local version of \texttt{cmake}.} However, the installation can also be performed independently at any time in the \maddm command line:
\begin{Verbatim}[fontsize=\fontsize{9.5}{11}\selectfont]
  MG5_aMC_v2_9_4$ python2.7 bin/maddm.py
  MadDM> install looptools
  MadDM> quit
  MG5_aMC_v2_9_4$
\end{Verbatim}
The computation of loop processes for indirect detection of dark matter using \maddm calls all the above external packages. Hence they should be appropriately cited. 
Now everything is set for the user to start computing loops and lines for her/his favorite dark-matter models.

\subsection{Generate command}

Once the user has entered the \maddm command line by executing "\texttt{python2.7 bin/maddm.py}", he/she can navigate through the program via the following commands. First, the user has to load a NLO UFO model. In the following, we will consider the model \linebreak\texttt{topphilic\_NLO\_EW\_CM\_UFO} as our reference benchmark model. This model is already available in the model repository of \mg and is automatically downloaded, when typing the appropriate command:
\begin{Verbatim}[fontsize=\fontsize{9.5}{11}\selectfont]
  MG5_aMC_v2_9_4$ python2.7 ./bin/maddm.py
  MadDM> import model topphilic_NLO_EW_CM_UFO
  MadDM> define darkmatter chi
\end{Verbatim}
where we have specified the dark-matter candidate (\texttt{chi}) in the last line. The phenomenology of this model has been considered in Sec.~\ref{sec:stopmodel}. It has only two BSM particles. Besides dark matter, it contains a coloured scalar mediator called \texttt{t1}. This model is a simplified version of the more general \texttt{DMSimpt} framework~\cite{Arina:2020udz,Arina:2020tuw} which describes very generic $t$-channel models coupling to all standard-model quarks.

With the current version, \maddm has the new feature enabling the automated recognition of all BSM particles that are odd under the $Z_2$-symmetry that stabilizes dark matter. For instance, for the model under consideration, after performing the \texttt{define darkmatter} command, \maddm displays:
\begin{Verbatim}[fontsize=\fontsize{9.5}{11}\selectfont]
  display z2-odd
  INFO: z2-odd: chi, t1 
  ...
\end{Verbatim}

The gamma-line observables introduced with this release are computed via
\begin{Verbatim}[fontsize=\fontsize{9.5}{11}\selectfont]
  MadDM> generate indirect_spectral_features
  MadDM> output my_process_dir
\end{Verbatim}
The \texttt{indirect\_spectral\_features}  command performs the automatic generation of all diagrams for the final states $\gamma\gamma$ and $\gamma X$, where  $X$ includes the standard-model $Z$ and $h$ as well as all additional BSM particles, that are kinematically accessible and even under the dark symmetry.
Hence, for the model under consideration the generated final states are $\gamma\gamma$, $\gamma Z$ and $\gamma h$ only, as \texttt{t1} is $Z_2$-odd. 

\begin{figure*}[t]
    \centering
    {
    \begin{minipage}{6cm}
        \dirtree{%
        .1 MG5\_aMC\_v2\_9\_4. 
        .2 bin. 
        .3 maddm.py.
        .3 mg5\_aMC.
        .3 \dots.
        .2 models.
        .2 my\_process\_dir.
        .3 bin.
        .3 Cards.
        .4 maddm\_card.dat.
        .4 multinest\_card.dat.
        .4 param\_card.dat.
        .4 \dots.
        .3 Indirect\_LI\_line.
        .4 Cards.
        .5 run\_card.dat.
        .5 MadLoopParams.dat.
        .5 \dots.
        .4 \dots.
        .3 output.
        .4 run\_01.
        .4 run\_02.
        .4 \dots.
        .3 \dots.
        .2 PLUGIN.
        .3 maddm.
        .3 \dots.
        .2 \dots.
        }
    \end{minipage}
        \begin{minipage}{8.5cm}
        \dirtree{%
        .1 run\_01.
        .2 maddm\_card.dat.
        .2 MadDM\_results.txt.
        .2 maddm.out.
        .2 Output\_Indirect\_LI\_line.
        .3 run\_01\_DM\_banner.txt.     
        .3 rwgt\_events\_rwgt\_1.lhe.gz.
        .3 unweighted\_events.lhe.gz.
        }
    \end{minipage}
    }
    \caption{Schematic structure of the folders and files of \mg and \maddm. {\bfseries{Left:}} The main directory of \mg. The python executable file \texttt{maddm.py} is located in the \texttt{bin} folder, while the source code is in \texttt{PLUGIN/maddm/}. The output directory \texttt{my\_process\_dir} contains all relevant setting cards (within \texttt{Cards}), and the output files in \texttt{output/run\_01} for instance. {\bfseries{Right:}} Zoomed view of the \texttt{run\_01} directory, where the main results are stored, as labeled. The file \texttt{MadDM\_results.txt} recaps the value of all observables computed by the user. The \texttt{Output\_Indirect\_LI\_line} contains the indirect detection lhe event files and is a symbolic link to the actual directory \texttt{Indirect\_LI\_line}. The latter directory contains the \texttt{MadLoopParams.dat} card to tweak the \ml settings if necessary, before executing the \texttt{launch} command.}
    \label{fig:folder}
\end{figure*}
The command \texttt{output} creates all the necessary code to perform the loop-induced computations. The above commands create the folder structure illustrated in figure~\ref{fig:folder}. The indirect detection directory called \linebreak\texttt{Indirect\_LI\_line} is created, where \texttt{LI} and \texttt{line} stand for loop-induced process and line signal, respectively. Inside this directory, in the \texttt{Cards} folder, all cards for settings are stored including the \texttt{MadLoopParams.dat} card, which serves to set the \ml run parameters. The output folder inside \texttt{my\_process\_dir} contains a sub-directory for each run performed for the generated process (\texttt{run\_01}, \texttt{run\_02}, \dots) which, in turn, contains all outputs of the computation (stored in \linebreak\texttt{MadDM\_results.txt} and \texttt{maddm.out}) as well as a copy of the \texttt{maddm\_card.dat} used. Further output is linked to the directory \texttt{Output\_Indirect}, see the next section for more details.

Instead of computing all possible channels that provide gamma-lines signatures at once the user may want to compute individual channels. This is possible by specifying the final state as an argument of the command. For instance, for the $\gamma\gamma$ final state, the syntax is
\begin{Verbatim}[fontsize=\fontsize{9.5}{11}\selectfont]
  MadDM> ...
  MadDM> generate indirect_spectral_features a a
  MadDM> output my_process_dir_aa
\end{Verbatim}
(Note that the photon is denoted by "\texttt{a}" here.)
According to the general \maddm syntax (see \emph{e.g.}~\cite{Arina:2020kko}) a specific subset of channels could be defined, for instance:
\begin{Verbatim}[fontsize=\fontsize{9.5}{11}\selectfont]
  MadDM> ...
  MadDM> generate indirect_spectral_features a a
  MadDM> add indirect_spectral_features a z
  MadDM> output my_process_dir_aa_az
\end{Verbatim}
As the computation of loop-induced process can be expensive (in particular for sophisticated models) we 
recommend specifying the final state if the computation of all channels is not needed.

\subsection{Launch command}\label{sec:launch}

\begin{figure*}[t!]
\centering
\begin{Verbatim}[fontsize=\scriptsize]
The following switches determine which programs are run:
/======================= Description =======================|========= values ==========|======== other options ========\
| 1. Compute the Relic Density                              |         relic = OFF       |     ON                        |
| 2. Compute direct(ional) detection                        |        direct = OFF       |     Please install module     |
| 3. Compute indirect detection/flux (cont spectrum)        |      indirect = OFF       |     Please install module     |
| 4. Compute indirect detection in a X (line spectrum)      |      spectral = ON        |     OFF                       |
| 5. Run Multinest scan                                     |      nestscan = OFF       |     ON                        |
\=======================================================================================================================/
 You can also edit the various input card:
 * Enter the name/number to open the editor
 * Enter a path to a file to replace the card
 * Enter set NAME value to change any parameter to the requested value
 /=============================================================================\ 
 |  6. Edit the model parameters    [param]                                    |  
 |  7. Edit the MadDM options       [maddm]                                    |
 \=============================================================================/
 [60s to answer]
 >
\end{Verbatim}
\vspace{-2.3ex}
\caption{Example of the launch interface after performing the \texttt{launch} command in \maddm in gamma-ray line computations. The indirect detection spectral feature calculations are turned on, while relic density, direct detection, continuum indirect detection and the scan capability with \textsc{MultiNest} are switched off.
}
\label{fig:prompt}
\end{figure*}

The loop-induced computations for a given parameter point are performed via the \texttt{launch} command:
\begin{Verbatim}[fontsize=\fontsize{9.5}{11}\selectfont]
  MadDM> launch my_process_dir
\end{Verbatim}
This opens the \emph{launch interface} that allows the user to change settings and model parameters as it is shown in figure~\ref{fig:prompt}. As in previous releases of \maddm, these changes can be made in two ways. First, (repeatedly) entering a number 1--5 allows to alternate between the options displayed: here we focus on item 4 ( indirect detection of spectral lines) which is the only one turned on in figure~\ref{fig:prompt}. Furthermore, entering 6 or 7 opens the files \texttt{param\_card.dat} or \texttt{maddm\_card.dat}, respectively, with a command-line editor (\texttt{vim} by default) to modify the settings. The former file contains all model parameters, while the latter allows the user to change most of the \maddm\ settings. Many new parameters related to the gamma-line analysis have been introduced with this release. They are described below. Alternatively, the user can enter the command \texttt{\,set~<parameter>~<value>\,} in the launch interface to control the various settings. For our considered benchmark model, we choose the following model parameters as an example:
\begin{Verbatim}[fontsize=\fontsize{9.5}{11}\selectfont]
  > set mchi 450.
  > set mst 622.
  > set lambdachi 0.85
\end{Verbatim}
These changes are stored in \texttt{param\_card.dat}.  Additionally, the astrophysical parameters for the gamma-ray line analysis can be specified, for instance:
\begin{Verbatim}[fontsize=\fontsize{9.5}{11}\selectfont]
  > set vave_indirect_line 7.5e-4
  > set profile einasto
  > set alpha 0.17
  > set r_s 20.0
  > set nevents 10000 
\end{Verbatim}
These settings are stored in \texttt{maddm\_card.dat}.
In the first line, the mean average velocity (in units of $c$) relevant for the considered gamma-ray line observations is set. For the applicability of the Fermi-LAT and HESS constraints implemented in \maddm, a value for the center of the Galaxy has to be entered. Accordingly, if \texttt{vave\_indirect\_line} lies outside $[5\times 10^{-4}; 10^{-3}]c$, a warning is displayed. Here we choose the corresponding value for the center of the Galaxy from~\cite{Eilers:2019} which is chosen by default.
Note that \texttt{vave\_indirect\_line} is independent of \texttt{vave\_indirect\_cont} the latter of which being the mean average velocity relevant for the computation of continuum spectra (\emph{e.g.}~for gamma-ray observations in dwarf spheroidal galaxies).\footnote{In previous versions of \maddm, the variable \texttt{vave\_indirect} set the average velocity relevant for the computation of continuum spectra. For backward compatibility, \texttt{vave\_indirect} can still be used.}

In the second to fourth lines, the dark-matter density profile is set to Einasto, with $\alpha=0.17$ and scale radius of 20\,kpc. These are the default parameters. These parameters are in accordance with the one chosen for the cross-section limit-setting in the Fermi-LAT analysis for the R16 ROI. In the last line, the number of generated events is set to 10000, which is the recommended ball-park to obtain a precise determination of the line annihilation cross-sections.

Note that the content of the file \texttt{maddm\_card.dat} can be adjusted according to the level of detail to which the user needs to control the settings. While the most common settings are contained in \texttt{maddm\_card.dat} by default (in particular the settings mentioned above) the user can switch to the full mode in which an extended set of parameters is displayed that are of interest for specific applications only. This can be achieved by typing
\begin{Verbatim}[fontsize=\fontsize{9.5}{11}\selectfont]
  MadDM> update to_full
\end{Verbatim}
directly after entering the launch interface. Note that the \texttt{maddm\_card.dat} file can be opened in a command-line editor by typing 7. Examples of parameters that can only be changed in the full mode appear in appendix~\ref{sec:app1}. An example for the full content of the \texttt{maddm\_card.dat} associated with gamma-line signatures can be found in figure~\ref{fig:maddm_card}.

Once the user has completed the settings, the run is launched by pressing enter. The launch interface is  closed and the results are displayed on screen. The screen output is shown in figure~\ref{fig:results}. It contains the following information:
\begin{figure*}[t!]
\centering
\begin{Verbatim}[fontsize=\scriptsize]
INFO: MadDM Results 
INFO: 
****** Indirect detection: 
INFO: <sigma v> method: madevent  
INFO: ====== line spectrum final states 
INFO: DM particle halo velocity: 0.00075 c  
INFO: *** Print <sigma v> [cm^3 s^-1] with Fermi-LAT 2015, HESS 2018 line limits 
INFO: chi chi > a h       All DM = 2.00e-40       ALLOWED   	Line GC ul       = 1.14e-27 
INFO: chi chi > a z       All DM = 2.58e-35       ALLOWED   	Line GC ul       = 1.11e-27 
INFO: chi chi > a a       All DM = 1.08e-34       ALLOWED   	Line GC ul       = 5.38e-28 
INFO:  
INFO: *** Line limits from Fermi-LAT 2015, HESS 2018 
INFO: Density profile: Einasto(rho_s = 8.1351e-02 GeV cm^-3, r_s = 2.00e+01 kpc, alpha = 1.70e-01) 
INFO: ==== Fermi-LAT 2015 ========================================================================================== 
INFO: ROI: 16.0 
INFO: J = 9.390000e+22 GeV^2 cm^-5 
INFO: detection range: 2.1400e-01 -- 4.6200e+02 GeV 
INFO: ------------------------------------------------------------------ 
INFO:                                          Flux [cm^-2 s^-1]         
INFO:                    Energy [GeV]   -------------------------------- 
INFO:                                     All DM                  UL     
INFO: ------------------------------------------------------------------ 
INFO: peak_1(ah+az+aa)    4.4958e+02    4.4734e-18 ALLOWED    5.1290e-11 
INFO: ------------------------------------------------------------------ 
INFO: ==== HESS 2018 =============================================================================================== 
INFO: ROI: 1.0 
INFO: J = 4.660000e+21 GeV^2 cm^-5 
INFO: detection range: 3.0690e+02 -- 6.3850e+04 GeV 
INFO: ------------------------------------------------------------------ 
INFO:                                          Flux [cm^-2 s^-1]         
INFO:                    Energy [GeV]   -------------------------------- 
INFO:                                     All DM                  UL     
INFO: ------------------------------------------------------------------ 
INFO: peak_1(ah+az+aa)    4.4951e+02    2.2200e-19 ALLOWED    9.9329e-13 
INFO: ------------------------------------------------------------------ 
INFO:  
INFO: Results written in: ~/MG5_aMC_v2_9_4/my_process_dir/output/run_01/MadDM_results.txt 
\end{Verbatim}
\vspace{-2.3ex}
\caption{Example of output provided by \maddm in gamma-ray line computations after the successful run of \ml. The numbers quoted correspond to the benchmark model point  taken as an example in~appendix~\ref{sec:launch}.
}
\label{fig:results}
\end{figure*}
\begin{itemize}
\item The annihilation cross-section has been computed with the reshuffling method. Note that the \maddm \texttt{fast mode} is disabled for loop-induced processes unless the loop process is described by an effective vertex, in which case the process arises at tree level; 
\item \maddm displays the velocity at which the processes is computed (as set above);
\item \maddm prints the value of the annihilation cross-section for the \texttt{aa}, \texttt{az} and \texttt{ah} separately, for the scenario called \texttt{All DM}. The latter  assumes that the dark-matter particles make up 100\% of the required relic abundance, being agnostic on how this is achieved (see~\cite{Arina:2020kko} for more details);
\item \maddm displays the corresponding experimental 95\% CL upper limits for the selected ROI and density profile (here R16 and Einasto, respectively). This information is only provided if the user chooses a density profile and ROI that match the experimental ones. Note that this is the case for the default settings. Here the model point is flagged \texttt{ALLOWED} as the theoretical predictions are below experimental bounds. It would be flagged \texttt{EXCLUDED} otherwise. 
Note that for model points that can be both constrained by Fermi-LAT and HESS, \maddm shows the most constraining limit among the two.
\item Below the line \,\texttt{INFO: *** Line limits from} \linebreak\texttt{Fermi-LAT 2015, HESS 2018},\, \maddm displays \linebreak the results associated with the flux computation. 
First, \maddm provides information about the dark-matter density profile. Below the results for the \linebreak Fermi-LAT and HESS analyses, respectively, are \linebreak shown. Note that results are only shown if $E_\gamma$ of at least one peak lies within the experimental range of the corresponding experiment ($\pm$ half of full width at half maximum). This is the case for both experiments for the chosen benchmark model. For each experiment, \maddm prints the $J$-factor for the chosen density profile and the corresponding ROI (see~appendix~\ref{sec:app1a} for details) and shows the energy range of the experiment. Finally, it displays the predicted flux and the corresponding 95\% CL flux upper limits together with the flag \texttt{ALLOWED} (or \texttt{EXCLUDED}). For the considered parameter point, the spectral lines for the three final states appear at the same energy (within the resolution of both experiments). Hence, they have been merged to a single one (see~appendix~\ref{sec:app3} for details). In the case of several distinguishable peaks, the corresponding results for each peak would be listed.
Note that due to the combination of channels the constraints from the flux computation can be stronger.

\end{itemize}
The same output information is stored in the file \linebreak\texttt{my\_process\_dir/output/run\_01/MadDM\_results.txt}, \linebreak see figure~\ref{fig:folder}.

\subsection{A complete example}\label{sec:full}

\begin{figure*}[t!]
    \centering
    {
    \begin{minipage}{6cm}
        \dirtree{%
        .1 my\_process\_dir\_complete.
        .2 Cards. 
        .2 output.
  	.2 Indirect\_LI\_cont.
	.3 Cards.
	.4 maddm\_card.dat.
        .4 run\_card.dat.
        .4 param\_card.dat.
        .4 MadLoopParams.dat.
        .4 \dots.
        .2 Indirect\_LI\_line.
        .3 Cards.
        .4 maddm\_card.dat.
        .4 run\_card.dat.
        .4 param\_card.dat.
        .4 MadLoopParams.dat.
        .4 \dots.
        .2 Indirect\_tree\_cont.
        .3 Cards.
        .4 maddm\_card.dat.
        .4 run\_card.dat.
        .4 param\_card.dat.        
        .4 \dots.
        .2 \dots.
        }
    \end{minipage}
     \begin{minipage}{8.5cm}
        \dirtree{%
        .1 output.
        .2 run\_01.
        .3 maddm\_card.dat.
        .3 MadDM\_results.txt.
        .3 maddm.out.
        .3 antiprotons\_spectrum\_pythia8.dat.
        .3 gammas\_spectrum\_pythia8.dat.
        .3 neutrinos\_e\_spectrum\_pythia8.dat.
        .3 neutrinos\_mu\_spectrum\_pythia8.dat.
        .3 neutrinos\_tau\_spectrum\_pythia8.dat.
        .3 positrons\_spectrum\_pythia8.dat.
        .3 Output\_Indirect\_LI\_cont.
        .4 run\_01\_DM\_banner.txt.     
        .4 rwgt\_events\_rwgt\_1.lhe.gz.
        .4 unweighted\_events.lhe.gz.
        .4 antiprotons\_spectrum\_pythia8.dat.
        .4 gammas\_spectrum\_pythia8.dat.
        .4 neutrinos\_e\_spectrum\_pythia8.dat.
        .4 neutrinos\_mu\_spectrum\_pythia8.dat.
        .4 neutrinos\_tau\_spectrum\_pythia8.dat.
        .4 positrons\_spectrum\_pythia8.dat.
        .4 \dots.
        .3 Output\_Indirect\_LI\_line.
        .4 run\_01\_DM\_banner.txt.     
        .4 rwgt\_events\_rwgt\_1.lhe.gz.
        .4 unweighted\_events.lhe.gz.
        .3 Output\_Indirect\_tree\_cont.
        .4 run\_01\_DM\_banner.txt.     
        .4 rwgt\_events\_rwgt\_1.lhe.gz.
        .4 unweighted\_events.lhe.gz.
        .4 antiprotons\_spectrum\_pythia8.dat.
        .4 gammas\_spectrum\_pythia8.dat.
        .4 neutrinos\_e\_spectrum\_pythia8.dat.
        .4 neutrinos\_mu\_spectrum\_pythia8.dat.
        .4 neutrinos\_tau\_spectrum\_pythia8.dat.
        .4 positrons\_spectrum\_pythia8.dat.
        .4 \dots.
        .3 \dots.
        }
    \end{minipage}
    }
    \caption{Example of the indirect detection folders and files of \maddm. {\bfseries{Left:}}  Structure of the user folder \texttt{my\_process\_dir\_complete} that contains all relevant setting cards inside the individual indirect folders (\texttt{Indirect\_LI\_line}, \texttt{Indirect\_LI\_cont} and \texttt{Indirect\_tree\_cont}), and the output files in \texttt{output/}. {\bfseries{Right:}} Zoomed view of the \texttt{output} directory, where the main results are stored. The file \texttt{MadDM\_results.txt} recaps the value of all observables computed by the user. The \texttt{Output\_Indirect\_LI\_cont}, \texttt{Output\_Indirect\_LI\_line} and \texttt{Output\_Indirect\_tree\_cont} contain the indirect detection lhe event files for loop-induced processes $gg$, $aX$ and tree level, respectively. 
    }
    \label{fig:totfolder}
\end{figure*}

While we concentrated on the gamma-line features in the previous examples we finally display a complete example involving the computation of the freeze-out relic density, of continuum fluxes (at tree level and loop-induced) and gamma-line fluxes.
The corresponding commands are:
 \begin{Verbatim}[fontsize=\fontsize{9.5}{11}\selectfont]
  MG5_aMC_v2_9_4$ python2.7 ./bin/maddm.py
  MadDM> import model topphilic_NLO_EW_CM_UFO
  MadDM> define darkmatter chi
  MadDM> generate relic_density
  MadDM> add indirect_detection
  MadDM> add indirect_detection g g
  MadDM> add indirect_spectral_features
  MadDM> output my_process_dir_complete
\end{Verbatim}
Here, we perform the relic density computation without specifying a potential coannihilator, because we consider a benchmark point in which \texttt{chi} and \texttt{t1} have a relatively large mass splitting rendering coannihilation to be irrelevant. In general, coannihilators should be considered, see~\cite{Arina:2020kko} for details. 
The first command \texttt{add indirect\_detection} generates the code necessary to perform tree-level indirect detection computations. The second command \texttt{add indirect\_detection g\;g}\, generates the code to compute the process into pairs of gluons, which is loop-induced but produces a continuum energy spectrum of cosmic messengers. All continuum spectra are added and the ones for gamma rays are automatically confronted with limits from Fermi-LAT observations of dwarf spheroidal galaxies. The third command \texttt{add indirect\_spectral\_features} adds the computation of spectral lines as discussed above. The combination of these three commands represents all possible ways of generating gamma rays from the annihilation of dark matter for this specific top-philic model.
The execution of these commands creates the folder structure displayed in figure~\ref{fig:totfolder}, inside the directory \linebreak\texttt{my\_process\_dir\_complete}.
In the example above, there are the following three folders corresponding to the three classes of processes generated:
\begin{enumerate}
    \item \texttt{Indirect\_tree\_cont}: this folder contains the tree-level indirect detection processes leading to continuum spectra of cosmic messengers. These processes are run with the dark-matter halo velocity \linebreak\texttt{vave\_indirect\_cont};
    \item \texttt{Indirect\_LI\_line}: this folder contains the gamma-ray line loop-induced processes. This run is pursued with the velocity  \texttt{vave\_indirect\_line};
      \item \texttt{Indirect\_LI\_cont} contains loop-induced processes leading to continuum spectra (here the \texttt{gg} final state). This run is pursued with  \texttt{vave\_indirect\_cont}. The photon energy spectrum generated from \texttt{gg} is merged together with the gamma-ray spectrum used for the continuum photon analysis based on the Fermi-LAT data. 
      As in previous versions the spectra can be generated using \py or be taken from \linebreak\pppc (in the case of standard-model final states).
\end{enumerate} 
Each indirect detection folder is run separately by \linebreak\maddm, one after the other. The different relevant settings  are contained in the \texttt{Cards} folder of each indirect detection directory. The user can change settings in the launch interface, as described below. 
       
Note that there is also a fourth possibility if the annihilation into photons proceeds via an effective vertex in the model.  
In this case, \maddm generates a directory  \texttt{Indirect\_tree\_line} that indicates that the process occurs at tree level but provides a line signature processed by the corresponding analysis pipeline. We will not comment on this case further as the behavior is similar to the example being discussed. Note, however, that in this case, the \maddm \texttt{fast} mode is available.

After performing the command
\begin{Verbatim}[fontsize=\fontsize{9.5}{11}\selectfont]
  MadDM> launch my_process_dir_complete
\end{Verbatim}
the \maddm interface is displayed, as shown in figure~\ref{fig:complete}. The relic density, indirect detection observables for continuum emission and line signatures are turned on, while the other modules are off. For the continuum emission module the mode \texttt{flux\_earth} is chosen. Other options are \texttt{flux\_source} and \texttt{sigmav}. For more information about these options we refer the reader to \cite{Arina:2020kko,Ambrogi:2018jqj}.
\begin{figure*}[t!]
\centering
\begin{Verbatim}[fontsize=\scriptsize]
The following switches determine which programs are run:
/===================== Description =====================|========= values ==========|========== other options ===========\
| 1. Compute the Relic Density                          |      relic = ON           |     OFF                            |
| 2. Compute direct(ional) detection                    |     direct = OFF          |     Please install module          |
| 3. Compute indirect detection/flux (cont spectrum)    |   indirect = flux_source  |     flux_earth|OFF|sigmav          |
| 4. Compute indirect detection in a X (line spectrum)  |   spectral = ON           |     OFF                            |
| 5. Run Multinest scan                                 |   nestscan = OFF          |     ON                             |
\========================================================================================================================/
 You can also edit the various input card:
 * Enter the name/number to open the editor
 * Enter a path to a file to replace the card
 * Enter set NAME value to change any parameter to the requested value
 /=============================================================================\ 
 |  6. Edit the model parameters    [param]                                    |  
 |  7. Edit the MadDM options       [maddm]                                    |
 \=============================================================================/
>
INFO: Start computing spectral,relic,indirect 
\end{Verbatim}
\vspace{-2.3ex}
\caption{Example of the launch interface after performing the \texttt{launch} command in \maddm for both tree level and loop-induced computations. The relic density, standard indirect detection and indirect detection spectral feature calculations are turned on, while direct detection and the scan capability with \textsc{MultiNest} are switched off.
}
\label{fig:complete}
\end{figure*}
Settings for the run are done by typing
\begin{Verbatim}[fontsize=\fontsize{9.5}{11}\selectfont]
  > set mchi 450.
  > set mst 622.
  > set lambdachi 0.85
  > set vave_indirect_cont 2.e-5
  > set vave_indirect_line 7.5e-4
  > set profile einasto
  > set alpha 0.17
  > set r_s 20.0
  > set nevents 10000 
\end{Verbatim}
These are almost the same settings described in~appendix~\ref{sec:launch}, the one addition concerns the dark-matter velocity for the computation of continuum fluxes (\texttt{vave\_indirect\_cont}), which is set to a value compatible with the one measured in dwarf spheroidal galaxies. The run is performed in the \maddm \texttt{precise} mode. The number of generated events, \texttt{nevents}, is common to all indirect detection observables, \emph{i.e.}~for each class of processes, 10000 events are generated. Note that this choice should be adjusted to the user's needs. For the computation of smooth energy spectra for the continuum emission (in particular in the tails) a significantly larger number of events may be required.

The final output screen of \maddm is displayed in figure~\ref{fig:outputtot}, and is constituted by three main blocks:
\begin{figure*}[t!]
\centering
\begin{Verbatim}[fontsize=\scriptsize]
INFO: MadDM Results 
INFO: Define xsi = Relic density/Planck measurement for thermal scenarios. 
INFO: Rescaling theory prediction for xsi(direct det.) and xsi^2(indirect det.) for thermal scenarios.
 
INFO: 
****** Relic Density
OMEGA IS 0.119894 
INFO: Relic Density       = 1.20e-01       WITHIN EXP ERROR   
INFO: x_f                 = 2.10e+01             
INFO: sigmav(xf)          = 1.14e-26 cm^3 s^-1   
INFO: xsi                 = 9.99e-01             
INFO:  
INFO: Channels contributions: 
INFO: chi chi > t t~      : 100.00 % 
INFO: 
****** Indirect detection: 
INFO: <sigma v> method: madevent  
INFO: ====== continuum spectrum final states 
INFO: DM particle halo velocity: 2e-05 c  
INFO: *** Print <sigma v> [cm^3 s^-1] with Fermi dSph limits 
INFO: chi chi > g g       Thermal = 7.95e-29   ALLOWED     All DM = 7.96e-29   ALLOWED    Fermi dSph ul  = 7.39e-26 
INFO: chi chi > t t~      Thermal = 1.55e-26   ALLOWED     All DM = 1.55e-26   ALLOWED    Fermi dSph ul  = 1.00e-25 
INFO: DM DM > all         Thermal = 1.56e-26   ALLOWED     All DM = 1.56e-26   ALLOWED    Fermi dSph ul  = 1.28e-25 
INFO:  
INFO: *** Fluxes at earth [particle cm^-2 sr^-1]: 
INFO: gammas Flux         =	3.49e-10             
INFO: neutrinos_e Flux    =	1.97e-13             
INFO: neutrinos_mu Flux   =	2.15e-13             
INFO: neutrinos_tau Flux  =	1.94e-13             
INFO:  
INFO: ====== line spectrum final states 
INFO: DM particle halo velocity: 0.00075 c  
INFO: *** Print <sigma v> [cm^3 s^-1] with Fermi-LAT 2015, HESS 2018 line limits 
INFO: chi chi > a a       Thermal = 5.66e-31   ALLOWED     All DM = 5.67e-31   ALLOWED    Line GC ul     = 5.38e-28 
INFO: chi chi > a z       Thermal = 1.34e-31   ALLOWED     All DM = 1.34e-31   ALLOWED    Line GC ul     = 1.11e-27 
INFO: chi chi > a h       Thermal = 1.04e-36   ALLOWED     All DM = 1.04e-36   ALLOWED    Line GC ul     = 1.14e-27 
INFO:  
INFO: *** Line limits from Fermi-LAT 2015, HESS 2018 
INFO: Density profile: Einasto(rho_s = 8.1351e-02 GeV cm^-3, r_s = 2.00e+01 kpc, alpha = 1.70e-01) 
INFO: ==== Fermi-LAT 2015 ========================================================================================== 
INFO: ROI: 16.0 
INFO: J = 9.390000e+22 GeV^2 cm^-5 
INFO: detection range: 2.1400e-01 -- 4.6200e+02 GeV 
INFO: ---------------------------------------------------------------------------------------- 
INFO:                                                     Flux [cm^-2 s^-1]                    
INFO:                    Energy [GeV]   ------------------------------------------------------ 
INFO:                                     All DM               Thermal                  UL     
INFO: ---------------------------------------------------------------------------------------- 
INFO: peak_1(ah+az+aa)    4.4958e+02    2.3397e-14 ALLOWED    2.3356e-14 ALLOWED    5.1290e-11 
INFO: ---------------------------------------------------------------------------------------- 
INFO: ==== HESS 2018 =============================================================================================== 
INFO: ROI: 1.0 
INFO: J = 4.660000e+21 GeV^2 cm^-5 
INFO: detection range: 3.0690e+02 -- 6.3850e+04 GeV 
INFO: ---------------------------------------------------------------------------------------- 
INFO:                                                     Flux [cm^-2 s^-1]                    
INFO:                    Energy [GeV]   ------------------------------------------------------ 
INFO:                                     All DM               Thermal                  UL     
INFO: ---------------------------------------------------------------------------------------- 
INFO: peak_1(ah+az+aa)    4.4951e+02    1.1612e-15 ALLOWED    1.1591e-15 ALLOWED    9.9328e-13 
INFO: ---------------------------------------------------------------------------------------- 
INFO:  
INFO: Results written in: ~/MG5_aMC_v2_9_4/my_process_dir_tot/output/run_01/MadDM_results.txt 
\end{Verbatim}
\vspace{-2.3ex}
\caption{Example of output provided by \maddm when running relic density and all indirect detection computations. The numbers quoted correspond to the benchmark model point  taken as an example.}
\label{fig:outputtot}
\end{figure*}
\begin{itemize}
\item The first block concerns the relic density.
The chosen benchmark point predicts the measured relic density within experimental errors, $\Omega h^2=0.12$. Furthermore, \maddm displays the temperature parameter at freeze-out $x_f$, the total thermally averaged annihilation cross section evaluated at the point $x_f$ and the dark-matter fraction of the considered candidate $\xi = \Omega h^2_{\rm theo}/ \Omega h^2_{\rm Planck}$ used for the rescaling of (in)direct detection limits. See~\cite{Arina:2020kko} for details. With the current version, \maddm also lists all annihilation channels and their contribution to the thermally averaged annihilation cross-section in percent. 
\item The second block shows the output for the (tree-level and loop-induced) processes leading to continuum fluxes. Since the relic density has been computed and the point is (slightly) under-abundant, the results are shown for both the \texttt{Thermal} and \texttt{All DM} scenario. In the former case, the limits are rescaled according to the dark-matter fraction $\xi$, while the latter assumes the measured relic density and hence no rescaling. For the benchmark point under consideration $\xi=0.99$, and hence no significant difference between the two cases. 

The corresponding exclusion limits at 95\% CL are given, for each final state separately. These are obtained by a simple  comparison with the tabulated exclusion limits in the \maddm experimental module. The line \texttt{DM DM > all} takes into account all computed channels (here $t\bar{t}$ and $gg$) and performs a likelihood analysis to provide the 95\% CL exclusion limit from the Fermi-LAT data from dwarf spheroidal galaxies. Note that to perform the likelihood analysis, the gamma-ray energy spectra inside \texttt{Output\_Indirect\_LI\_cont} are combined with those inside \texttt{Output\_Indirect\_tree\_cont}. Their sum is given directly in the folder \texttt{output/run\_01/}, as illustrated in figure~\ref{fig:totfolder}.
\item In the third block \maddm displays the annihilation cross sections and fluxes for the gamma-line signals, as already discussed in~appendix~\ref{sec:launch}. Here, however, since the relic density variable has been computed, both dark-matter scenarios, \linebreak\texttt{Thermal} and \texttt{All DM}, appear, see the previous bullet point for details. 
\end{itemize}

Finally, we recall that auto-completion is available to conveniently find commands and parameter names. Note further that all commands and settings described in the previous sections can be performed via a script whose path is passed as an argument with the execution of \texttt{maddm.py}, see~\cite{Arina:2020kko} for details.

\section{Astrophysics of gamma-ray lines}\label{sec:app1}

\subsection{\texorpdfstring{$J$}{J}-factor and dark-matter density profiles}\label{sec:app1a}

In this appendix, describe how the $J$-factor is computed within \maddm. 
By definition, eq.~\eqref{eq:jfactor}, the $J$-factor is:
\begin{equation}\label{eq:jfactor_app}
  J =  \int_{\rm ROI} {\rmd \Omega}\int_\text{los} \rho^2(\vec{r})\,  {\rmd}l\,.
 \end{equation}
Here we consider the ROIs that are used by the experimental searches. In general, they introduce masks such that the integration region in eq.~\eqref{eq:jfactor_app} goes beyond a simple cone.
The integrals above can be re-cast in the following form:
\begin{equation}
J = \int_A {\rm d^3} r^\prime \frac{1}{|\vec{r}^\prime|^2}\rho \left({\vec{r}(\vec{r}^\prime)}\right)^2 \,,
\label{eq:j-factor_app}
\end{equation}
where $A$ is the integration region and the integral is computed over $\vec{r}^\prime$, \emph{i.e.}~in the reference frame of the Sun, while the variable $r$ is in the reference frame of the Galactic center.
\begin{figure}[t!]
\centering
\includegraphics[width=.26\textwidth]{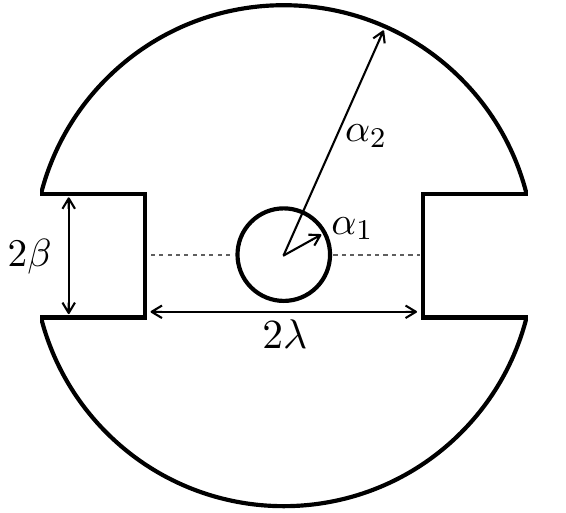}
\caption{2D projection at the Galactic center of the integration region $A$ defined in the $J$-factor integral. 
\label{fig:mask}}
\end{figure}
The integration region $A$ is defined as follows:
\begin{equation}
\begin{multlined}
A = \biggl\{\, (x,y,z) \;\Big\vert\;  x^2 + y^2 + z^2 \leq R^2\,  \cap \, \lvert y \rvert \leq x \tan (\lambda) \\
\cap \,   x^2 + y^2 \leq z^2 \tan^2 \biggl(\frac{\pi}{2} - \beta \biggr) \;\;\;\\
\cap \,  \lvert x \rvert \tan (\alpha_1 ) \leq \sqrt{z^2 + y^2} \leq \lvert x\rvert \tan (\alpha_2) \, \biggr\} \,,
\end{multlined}
\end{equation}
A two-dimensional projection of the integration region $A$ at the Galactic center is shown in figure~\ref{fig:mask}. It visualizes the parameters of the two possible masks considered in \maddm. On the one hand, it allows one to mask the Galactic plane up to a latitude $\beta$ and down to a Galactic longitude $\lambda$. On the other hand, it allows for a circular mask around the Galactic center with an opening angle $\alpha_1\in ( 0, \alpha_2)$. 
The outer opening angle of the ROI is denoted by $\alpha_2\in \left( 0, \pi \right)$. 

We use galactic coordinates for the computation: longitude $l \in \left(-\pi, \pi \right)$ and latitude $b \in \left( -\pi/2, \pi/2 \right)$.
For definiteness the galactic coordinates are:
\begin{subequations}
\begin{align}
x &= r^\prime \cos b \cos l \,, \\
y &= r^\prime \cos b \sin l \,, \\
z &= r^\prime \sin b \,.
\end{align}
\end{subequations}
Here, we assume that the integration cone is centered around the $x$ axis.

The mask parametrization introduced above allows one to describe a variety of ROIs, in particular, those considered by the Fermi-LAT and HESS collaborations. For instance, the R1 ROI used by the latter is obtained for $\alpha_1=0^\circ$, $\alpha_2=1^\circ$, $\lambda=0^\circ$ and $\beta=0.3^\circ$.
We have validated the computation of $J$-factors using the ones reported by the collaborations~\cite{Ackermann:2015lka,Abdallah:2018qtu} and elsewhere in the literature~\cite{Cirelli:2010xx} and find agreement below the percent level.\footnote{An exception is the $J$-factor for the NFWc density profile and R3 ROI~\cite{Ackermann:2015lka} where we find a 5\% deviation.}

We have included four dark-matter density profiles, specified in eqs.~\eqref{eq:nfwgen} (gNFW),~\eqref{eq:einasto} (Einasto),~\eqref{eq:burkert} (Burkert) and~\eqref{eq:iso} (isothermal). These dark-matter profiles are all normalized at the Sun position, the normalization is given by the value of the scale radius $r_s$ and by the scale density $\rho_s$. The latter is not a free parameter and is automatically computed once the user has fixed $\rho_\odot,\, R_\odot$ and $r_s$. 
The dark-matter density profile and the scale radius can be changed by the user by editing the file \texttt{maddm\_card.dat} or via the launch interface, for example:
\begin{Verbatim}[fontsize=\fontsize{9.5}{11}\selectfont]
  MadDM> set profile nfw
  MadDM> set r_s 25
\end{Verbatim}
These commands change the dark-matter density profile from its default Einasto to NFW with (which corresponds to gNFW with $\gamma=1$) and the scale radius to 25 kpc (the default value is $r_s=20$ kpc). 
The default values for the measured energy density at the Sun position and the Sun's distance from the Galctic center are $\rho_\odot = 0.4 \, \rm GeV/cm^3$ and $R_\odot=8.5 \, \rm kpc$, respectively, following the choice of the Fermi-LAT collaboration. Note that these parameters are not contained in the standard \texttt{maddm\_card.dat}. The change of these parameters requires enabling the full mode via typing \,\texttt{update\;to\_full} directly after entering the launch interface, see~appendix~\ref{sec:launch}. An example of the file \texttt{maddm\_card.dat} in full mode is given in figure~\ref{fig:maddm_card}. 
However, in the given example, \maddm computes the value of $\rho_s$ that corresponds to the default $\rho_\odot$, $R_\odot$  and to $r_s=25$ kpc. 

For NFWg, the inner slope parameter $\gamma$ is free. Users can change it by typing:
\begin{Verbatim}[fontsize=\fontsize{9.5}{11}\selectfont]
  MadDM> set profile gnfw
  MadDM> set gamma 1.3
\end{Verbatim}
This choice corresponds to NFWc density profile. 
Considering the Einasto, the slope $\alpha$ is a free parameter and can be set by
\begin{Verbatim}[fontsize=\fontsize{9.5}{11}\selectfont]
  MadDM> set profile einasto
  MadDM> set alpha 0.2
\end{Verbatim}
which change $\alpha$ from its default value $0.17$ to $0.2$.

\begin{figure*}[p]
\centering
\begin{Verbatim}[fontsize=\scriptsize]
# Method of computation (affects speed and precision)
   reshuffling = sigmav_method ! inclusive, madevent, reshuffling
   pythia8   = indirect_flux_source_method ! pythia8, PPPC4DMID, PPPC4DMID_ew
   dragon    =  indirect_flux_earth_method ! dragon, PPPC4DMID_ep

# setting for DM velocity
   2e-05     = vave_indirect_cont
   0.00075   = vave_indirect_line
   
# PPPC4DMID_ep settings for dm halo profile, propagation method and halo function
   Ein       = dm_profile ! NFW, Moo, Iso, Ein 
   MED       = prop_method ! MIN, MED, MAX
   MF1       = halo_funct ! MF1, MF2, MF3

# --------------------
# Line analysis
# --------------------

# choice of the dark matter density profile
   einasto   = profile ! gnfw, einasto, nfw, isothermal, burkert

# profile common parameters
   20.0      = r_s # kpc

# gnfw specific parameters
   1.3       = gamma

# einasto specific parameters
   0.17      = alpha

# profile normalisation parameters
   8.5       = r_sun # kpc
   0.4       = rho_sun # GeV cm^-3

# specify a ROI for Fermi-LAT 2015
# set default to use the ROI for which your profile is optimized
   default   = roi_fermi_2015 ! default, R3, R16, R41, R90

# number of FWHM to take as the minimum energy separation between peaks to consider them well separated
   5.0       = n_fwhm_separation

# minimum ratio between peaks' heights to have one of them significantly higher than the other
   10.0      = peak_height_factor

# toggle line experiments on/off
   on        = toggle_fermi_2015
   on        = toggle_hess_2018

### Template line experiment
# in the following you are able to specify the parameters for an experiment of your choice
# - toggle it on/off
   off            = toggle_template_line_experiment
# - name and arxiv (only for reference purpose)
   TEMPLATE_name  = template_line_experiment_name
   TEMPLATE_arxiv = template_line_experiment_arxiv
# - you can specify up to 1 ROI (in deg) and a profile associated to it (with the various parameters as seen above)
   1.0            = template_line_experiment_roi
   einasto        = template_line_experiment_profile
   20.0           = template_line_experiment_r_s
#  - gnfw parameter gamma
   1.3            = template_line_experiment_gamma
#  - einasto parameter alpha
   0.17           = template_line_experiment_alpha
# - specify the parameters of the mask (angles are in deg), refer to MadDM v3.2 documentation
   0.0            = template_line_experiment_mask_latitude
   180.0          = template_line_experiment_mask_longitude
   0.0            = template_line_experiment_mask_inner_angle
# - energy resolution: this is a percent value to be multiplied by the energy of the peak
   10.0           = template_line_experiment_energy_resolution
# - detection range in GeV
   0.0            = template_line_experiment_detection_range_min
   1000000.0      = template_line_experiment_detection_range_max
# - constraints file name: to be placed in $MADDM_PATH/ExpData/, must be 3 columns
#   DM mass (GeV), <sigmav> (cm^3 s^-1), flux (cm^-2 s^-1)
   None           = template_line_experiment_constraints_file
\end{Verbatim}
\vspace{-2.5ex}
\caption{Example of the content of \texttt{maddm\_card.dat} associated with gamma-line signatures including the extended set of parameters (full mode).}
\label{fig:maddm_card}
\end{figure*}

\subsection{Details about experimental data}\label{sec:app1b}

In general, the user can set the ROI for a given experiment.
Concretely, the experimental bounds are derived for a ROI which is optimized for a certain dark-matter density profile. 
In the case of Fermi-LAT there are four ROIs:
\begin{enumerate}
\item R3 associated with NFWc
\item R16 associated with Einasto
\item R41 associated with NFW
\item R90 associated with isothermal
\end{enumerate}
For HESS, only one ROI is defined, namely R1.
While the user is free to choose any of the four ROIs of the Fermi-LAT analysis by the variable \verb!roi_fermi_2015! in the file \verb!maddm_card.dat!, we recommend choosing the ROI associated with the dark-matter density profile considered. This can automatically be achieved by setting the above parameter to \verb!default!.\footnote{For the Burkert profile, the R90 is chosen by default.} Note that if a ROI is chosen that is not optimized for the considered profile, \maddm raises a warning.
If the user only chooses the ROI without specifying the dark-matter density profile, \maddm automatically chooses the profile for which it was optimized.
Note further that the approximate Fermi-LAT likelihood and $p$-value computations are only performed if the user chooses the R3 or R16 ROI. 

The experimental data is stored in the \linebreak directory \texttt{MG5\_aMC\_v2\_9\_4/PLUGIN/maddm/ExpData}. It \linebreak contains the respective data files:
\begin{Verbatim}
Fermi_2015_lines_R3_NFWcontracted.dat
Fermi_2015_lines_R16_Einasto.dat
Fermi_2015_lines_R41_NFW.dat
Fermi_2015_lines_R90_Isothermal.dat
\end{Verbatim}
for the Fermi-LAT analyses and 
\begin{Verbatim}
HESS_2018_lines_R1_Einasto.dat
\end{Verbatim}
for the HESS analysis. Each file contains the parameters of the density profile, the corresponding $J$-factor, the ROI and the 95\% CL upper limits on $\sigmav_{\gamma\gamma}$ and the flux in the 120 (60) energy bins considered by Fermi-LAT (HESS).

In addition to the implemented constraints from Fermi-LAT and HESS, we allow the user to consider further experimental data in the analysis pipeline. To this end, we have included a template experiment. It can, for instance, be used to compute the projected limits for future observations.
The associated parameters are among the extended set of parameters accessible in \verb!maddm_card.dat! in the full mode (see~appendix~\ref{sec:launch}). The respective block is shown in figure~\ref{fig:maddm_card} below the line \,\texttt{Template line experiment}.
The relevant parameters are:
\begin{itemize}
    \item \texttt{name}, \texttt{arxiv}: specify the name and (if applicable) the arXiv number of the analysis (optional);
    \item \texttt{roi}, \texttt{profile}: it is possible to specify the ROI (\emph{i.e.}~$\alpha_2$ in degrees) and the dark-matter density profile associated with it (the user can choose between the same profiles as for the main \texttt{profile} parameter discussed above);
    \item \texttt{r\_s}, \texttt{gamma}, \texttt{alpha}: these are the profile parameters to be specified for the default dark-matter density profile chosen as in the point above;
    \item \texttt{mask\_} parameters: allow the user to specify a mask, according to figure~\ref{fig:mask}, where \texttt{mask\_latitude} is $\lambda$, \texttt{mask\_longitude} is $\beta$ and \texttt{mask\_inner\_angle} is $\alpha_1$; they are all expressed in degrees;
    \item \texttt{energy\_resolution}: the relative energy resolution of the experiment in per-cent; taken to be constant over the detection range;
    \item \texttt{detection\_range\_} parameters: allow the user to specify the minimum (\texttt{min}) and maximum (\texttt{max}) energy range of the experiment (in GeV);
    \item \texttt{constraints\_file}: additionally, the user can provide the (projected) constraints of the experiment via a data file in the same format as \eg \linebreak\texttt{Fermi\_2015\_lines\_R90\_Isothermal.dat}. The file \linebreak should be included in the directory \linebreak\texttt{MG5\_aMC\_v2\_9\_4/PLUGIN/maddm/ExpData} of \linebreak\maddm. It should contain three columns: the dark-matter mass (in GeV), $\sigmav_{\gamma\gamma}$ upper limits (in \linebreak$\mathrm{cm}^{3}\,\mathrm{s}^{-1}$) and $\Phi_{\gamma\gamma}$ upper limits (in $\mathrm{cm}^{-2}\,\mathrm{s}^{-1}$).
\end{itemize}
To enable the template experiment, the parameter \linebreak\texttt{toggle\_template\_line\_experiment}
has be to set to \texttt{on}.
In this case, \maddm considers it in the same way as the constraints from Fermi-LAT and HESS.
Note that the dark-matter density profile associated with the template experiment is supposed to be the one for which the cross section upper limit has been derived (if provided by the user).
The density profile considered for the $J$-factor computation within \maddm remains the one specified through the parameter \texttt{profile} in \texttt{maddm\_card.dat}.
However, if the two profiles differ, a warning is raised.

Note that it is also possible to switch the Fermi-LAT and HESS analysis \texttt{on}/\texttt{off}.
When switched \texttt{off} the respective experiment is not considered in the analysis pipeline.

\section{Merging algorithm for peaks}\label{sec:app3}

In the presence of multiple spectral lines, the application of experimental constraints  (performed under the assumption of a single gamma-ray line signal) requires particular attention.
In the following, we provide details on the implemented merging algorithm for multiple spectral lines that are close in energy. Furthermore, we discuss the considered criteria for the applicability of the experimental analyses in this case.

Due to the non-relativistic nature of cold dark matter, the width of the gamma-ray line signal is expected to be very small. In particular, it is assumed to be small compared to the experimental energy resolution. In the experimental measurement, the sharp peak, eq.~\eqref{eq:line_spectrum_delta}, is hence smeared. We approximate it by a Gaussian,
\begin{equation}\label{eq:gausspeak}
\frac{\rmd\Phi_X^\textup{exp}(E)}{\rmd E} \propto \frac{\Phi_X}{\sigma \sqrt{2 \pi}} \exp{\left(-\frac{1}{2} \frac{\left( E - E_\gamma \right)^2}{\sigma^2} \right)}\,,
\end{equation}
where $\sigma$ corresponds to the experimental resolution, related to the full width at half maximum by $\rm FWHM = 2 \sqrt{2 \ln 2} \, \sigma$.
The resolution of Fermi-LAT is reported in~\cite{Ackermann:2015lka}. We
consider the EDISP3 energy resolution, associated with the best energy reconstruction estimator. It ranges from $\sim$ 10\% at 300 MeV down to 5.5\% at 500 GeV. 
Note that the width of each bin corresponds to the energy resolution (68\% containment) for that bin following EDISP3. The HESS telescope has a resolution in the energy of $10\%$ above $300$ GeV~\cite{Abdallah:2018qtu}. 

\begin{figure*}[tp]
\centering
\includegraphics[width=0.7\linewidth]{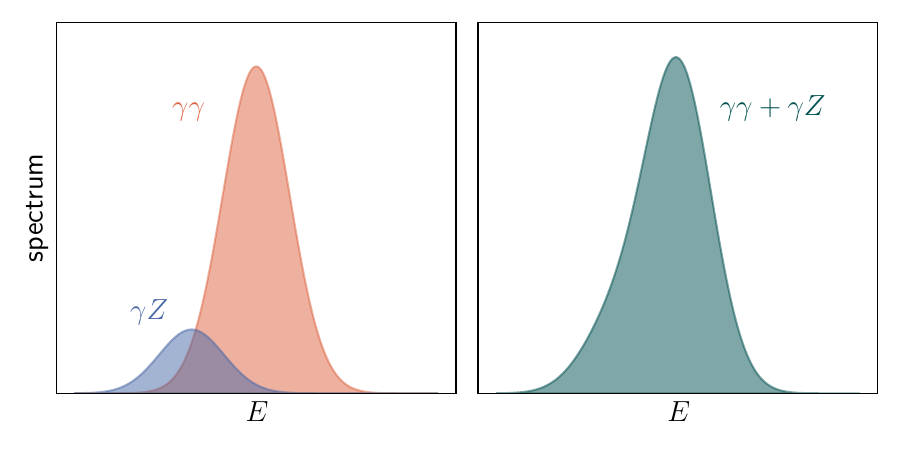}
\caption{Graphical example of the merging algorithm: on the left the $\gamma\gamma$ and $\gamma Z$ peaks are shown separately, while on the right the obtained merged peak is shown.}
\label{fig:peaksmerging}
\end{figure*}

The merging algorithm follows an iterative process. In each iteration, it considers all possible pairs of spectral lines (called peaks in the following) and selects the pair that is closest in energy, \emph{i.e.}~for which 
\begin{equation}
    s_{ab} = \frac{\lvert E_a - E_b \rvert}{\min\bigl(\mathrm{FWHM}_a,\mathrm{FWHM}_b\bigr)}\,,
    \label{eq:line_peaks_difference_fwhm_quantity_to_compute}
\end{equation}
is minimal, where $a,b$ index the peaks and ${\rm FWHM}_a$ is the FWHM of peak $a$. If $s_{ab}<1$, the Gaussian signals, eq.~\eqref{eq:gausspeak}, of the peaks are summed and considered as a single (merged) peak.
The iterative process is pursued with the set of merged and remaining un-merged peaks until no merging is possible anymore (under the above criterion). An illustrative example of a merging of two peaks is shown in figure~\ref{fig:peaksmerging}. 

The set of peaks obtained after merging represents the signal as potentially being observed by the experiment. 
However, if the signal still contains multiple peaks the application of the experimental constraints to each of them is, in general, questionable as the limit-setting procedure has been performed under the hypothesis of a single peak. Similar issues can arise if a merged peak has become too broad.
We address these concerns by introducing the following criteria the peaks have to satisfy. First, we require the FWHM of any merged peak to be smaller than 1.5 times the FWHM corresponding to the experimental resolution at its peak energy. If this criterion is not met, the peak is flagged with an error (error type 1).
Secondly, we require a minimal separation between the (merged) peaks. The minimal separation in units of the FWHM is controlled by the parameter \verb!n_fwhm_separation! whose default value is set to 5. 
Peaks that do not satisfy the separation criterion are flagged with the error type 2.
Note that the separation criterion is bypassed for a large hierarchy between the corresponding fluxes of the peaks, \emph{i.e.}~if the ratio of fluxes (each normalized to the upper limit at the corresponding energy) is larger than the variable \verb!peak_height_factor! (with default value 10). In this case, only the peak with the smaller flux is flagged with the error 2.

\bibliographystyle{spphys.bst} 
\bibliography{biblio}

\end{document}